\documentclass[pre,reprint,amsmath,floatfix,superscriptaddress,nofootinbib, aps, 10]{revtex4-1}

\usepackage[left=1.5cm, right=1.5cm, top=1.785cm, bottom=2.0cm]{geometry}

\usepackage[]{graphicx}
\usepackage[english]{babel}
\usepackage{array}

\usepackage[T1]{fontenc}
\usepackage{hyperref}
\usepackage{mathrsfs,amsmath}

\usepackage{epstopdf}

\pdfoutput=1

\begin{document}

\author{Sarah Trinschek}
\affiliation{Institut f\"ur Theoretische Physik, Westf\"alische Wilhelms-Universit\"at M\"unster, Wilhelm Klemm Str.\ 9, 48149 M\"unster, Germany}
\affiliation{Universit\'e Grenoble-Alpes, CNRS, Laboratoire Interdisciplinaire de Physique, 38000 Grenoble, France}

\author{Karin John}
\affiliation{Universit\'e Grenoble-Alpes, CNRS, Laboratoire Interdisciplinaire de Physique, 38000 Grenoble, France}

\author{Uwe Thiele}
\email{u.thiele@uni-muenster.de}
\affiliation{Institut f\"ur Theoretische Physik, Westf\"alische Wilhelms-Universit\"at M\"unster, Wilhelm Klemm Str.\ 9, 48149 M\"unster, Germany}
\affiliation{Center of Nonlinear Science (CeNoS), Westf{\"a}lische Wilhelms-Universit\"at M\"unster, Corrensstr.\ 2, 48149 M\"unster, Germany}
\affiliation{Center for Multiscale Theory and Computation (CMTC), Westf{\"a}lische Wilhelms-Universit\"at, Corrensstr.\ 40, 48149 M\"unster, Germany}

\title{Modelling of surfactant-driven front instabilities in spreading bacterial colonies}


\begin{abstract}
The spreading of bacterial colonies at solid-air interfaces 
is determined by the 
physico-chemical properties of the
involved interfaces. The production of surfactant molecules by bacteria is a widespread strategy that allows the
colony to efficiently expand over the
substrate. On the one hand, surfactant molecules lower the surface tension of the
colony, effectively increasing the wettability of the substrate, which facilitates spreading. On the other hand, 
gradients in the surface concentration of surfactant molecules result in
Marangoni flows that drive spreading. These flows may cause
an instability of the circular colony shape and the subsequent
formation of fingers.
In this work, we study the effect of bacterial surfactant production
and substrate wettability
on colony growth and shape within the
framework of a hydrodynamic thin film model.
We show that variations in the wettability and surfactant
production are sufficient to reproduce four different types of colony growth, which
have been described in the literature, namely, arrested and continuous spreading of circular colonies, slightly modulated front lines and the formation of pronounced fingers. 
\end{abstract}

\maketitle


\section{Introduction}
Bacteria are able to colonize solid-air interfaces by the formation of
dense colonies.\cite{Donl2002eid} After the attachment of individual
bacteria to the surface, they proliferate and a dense colony starts to
expand laterally over the surface. 
In many cases, the spreading is not driven by the active mobility of individual bacteria but rather by growth
processes and passive flows that result from the physico-chemical properties
of the bacterial film and the substrate.\cite{SSD+2002ARoM,YTS+2017bj}
One well studied example 
is the osmotic
spreading 
of biofilms, where the bacteria secrete an
extracellular matrix that acts as an osmolyte and triggers the influx of nutrient-rich water from
the underlying moist agar substrate into the colony, which
subsequently swells and spreads out.\cite{SAW+2012PNASUSA, YTS+2017bj,
  TJT2016ams, DTH2014prsb, YNS+2017nc} 

Another physical effect that plays a role in the spreading of bacterial
colonies at solid-air interfaces are wetting phenomena, which govern
the motion of the three-phase contact line between
the colony, the underlying agar substrate, and the surrounding
air.
For many bacterial strains, the molecules which are involved in the
quorum sensing mechanism (which allows for a cell-cell communication) have been found to play a double
role. Beside their signalling function, they act as bio-surfactants
(small molecules which adsorb to surfaces, thereby lowering the
surface tension) at
physiologically relevant concentrations.\cite{RR2001Em, RDN+2010Fmr}
Measurements of surface tension and contact angle
\cite{KHC+2015fm, LMB+2006AoM} indicate that 
bio-surfactants 
promote the spreading of bacterial colonies by improving wettability.
Additionally, gradients in surfactant concentration at the edges of the
colony give rise to so-called Marangoni fluxes which further drive 
cooperative spreading.\cite{FPB+2012sm,DFM+2015,YTS+2017bj,CSO2005job}  

For
\textit{Rhinozobium etli}, genetic knock-out
experiments\cite{DRH+2006pnasu} show that \textit{AHL}
(N-acyl-homoserine lactone) molecules are crucial for an efficient
swarming of the colony. The experimentally observed spreading speeds
and colony shapes are consistent with those estimated from a spreading
driven by Marangoni forces.  Growth measurements verify that for
\textit{Paenibacillus dendritiformis} colonies, the spreading velocity
indeed depends on the surfactant concentration but not on the
individual bacterial motion.\cite{BSZ+2009job} Genetic and
physico-chemical experiments \cite{FPB+2012sm, YTS+2017bj, KSF2003job,
  ARK+2009pnasu,CSO2005job} show that also in \textit{Bacillus subtilis} and
\textit{Pseudomonas aeruginosa} colonies, the surface tension gradient
induced by the respective bio-surfactants \textit{surfactin} and
\textit{rhamnolipids} is an important driver of colony
expansion. Further support of this theory comes from the demonstration
that swarming can be inhibited by the \textit{rhamnolipid} production
of nearby colonies \cite{TRL+2007em} as well as by the addition of
purified \textit{rhamnolipids} to the agar substrate \cite{CSO2005job}
as both suppress the necessary gradients in surface tension.
Besides enhancing the spreading speed, Marangoni fluxes may also be
responsible for the striking dendritic or finger-like colony patterns
observed in swarming experiments. 
Surfactant-producing \textit{Pseusomona aeruginosa}
colonies spread outwards and form pronounced fingers whereas a mutant
strain deficient in surfactant production can not expand and is
arrested in a small circular shape.\cite{CSO2005job,FPB+2012sm}

In the surfactant-assisted spreading of liquid drops (see
\onlinecite{MC2009sm} for a review), Marangoni fluxes are known to
give rise to a fingering front instability as first observed
experimentally in \onlinecite{ML1981cec} and subsequently confirmed
and studied in detail, e.g., in \onlinecite{TWS1989prl, HK1995pf,
  CCB+1999csa, ALM2003l, ALM2004l}. Numerical time simulations and
transient growth analysis of hydrodynamic thin-film models show
the presence of the instability for films 
covered by insoluble
surfactants \cite{THS1990prl, MT1999pof}, but are also extended
to soluble surfactants with sorption kinetics \cite{WCM2004pof} and
micelle formation.\cite{CM2006pof} 
In the context of
surfactant-mediated spreading of bacterial colonies, similar thin
film models are successfully applied to study the movement of a
\textit{Bacillus subtilis} biofilm up a wall on waves of surfactant\cite{ARK+2009pnasu} or
bacterial swarming in colonies of
\textit{Pseudomonas aeruginosa} in
a one-dimensional setting \cite{FPB+2012sm}. However,
two-dimensional hydrodynamic simulations which focus on the shapes of
spreading bacterial colonies driven by Marangoni effects have - to
the best of our knowledge - not yet been performed.  
Note that besides the surfactant-induced instability, also nutrient limitation and chemotactic effects are a possible causes 
for the dendritic morphology of bacterial colonies (for a critical review, see \onlinecite{MHH+2010mmonp}).
In this work, we
present a model for the surfactant-driven spreading of bacterial
colonies, which explicitly includes wetting effects. This allows us to
study the interplay between wettability and Marangoni fluxes and their
effect on the spreading speed and morphology. 
In section 2 we introduce the model, a passive thin-film
model with insoluble surfactants \cite{TAP2012pf} supplemented by
bioactive source terms. In section 3 we present a transversal linear
stability analysis to explore the possibility of front instabilities and
perform some full numerical simulations. 
\section{Thin film modelling of surfactant-driven biofilm spreading}
\label{sec:mod}
\begin{figure}
\centering
 \includegraphics[width=0.5\textwidth]{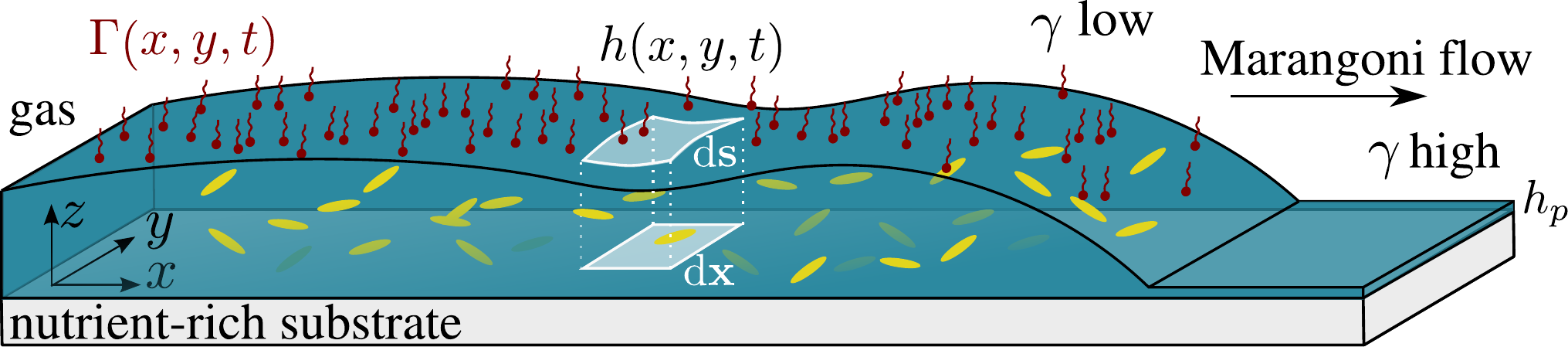}
 \caption{Sketch of a bacterial colony covered by an insoluble surfactant with concentration $\Gamma(x,y,t)$. The field $h(x,y,t)$ describes the height of liquid and biomass. The surfactant concentration on the colony is higher then on the surrounding substrate, resulting in gradients in the liquid-gas surface tension and outward-pointing Marangoni flows that promote the expansion of the colony.}
  \label{Fig:Biofilm_Surfactant_Sketch}
\end{figure}
A bacterial colony is a complex fluid, composed of water, bacteria,
nutrients and molecules, which are secreted by the bacteria,
e.g. extracellular polymeric substances  and surfactants.
In this work, we follow a simple two-field modelling strategy that
allows for a selective study of the influence of wettability and Marangoni
fluxes on the spreading dynamics. We treat the bacterial colony as a
thin film of height $h(x,y,t)$ covered by insoluble surfactant molecules of
concentration $\Gamma(x,y,t)$ [see
Fig.~\ref{Fig:Biofilm_Surfactant_Sketch}]. To model the
surfactant-driven spreading of the colony, we supplement the
hydrodynamic description with bioactive growth processes for the film
height $h$ and the surfactant concentration $\Gamma$. This approach is valid in the limit of fast osmotic equilibration
between the colony and the agar substrate.\cite{TJL+2017prl}
Similar models - which represent just one class in the very rich literature concerning the mathematical modelling of bacterial colonies (for reviews see for example \onlinecite{WZ2010ssc,KD2010sr,HL2014,PV2003}) - are used to study the influence of wettability \cite{TJT2016ams,TJL+2017prl}, quorum sensing \cite{WK2012jem,WKK+2001IJMAMB} and the surfactant-driven spreading of bacterial colonies.\cite{ARK+2009pnasu,FPB+2012sm} \\
In the following, we first present the 'passive' part of the hydrodynamic model for thin, surfactant-covered films before introducing the bioactive terms.
\subsection{Passive part of the model}
We consider a thin film of height $h(x,y,t)$ which is covered by an insoluble surfactant of (area-)density $\Gamma(x,y,t)$.
The description of the passive part of the model is based on the free energy functional
\begin{equation}
F[h,  \Gamma] = \int \left[ f_w(h)+ f_s \left( \Gamma \right) \zeta
\right] \mathrm{d \mathbf{x}}
\label{eq:energyfunctional}
\end{equation}
with
\begin{equation}
\zeta = \sqrt{1 + |\nabla h|^2} \approx 1 + \tfrac{1}{2} |\nabla h|^2 \, .
\label{eq:zeta}
\end{equation}
$F[h,  \Gamma]$ in (\ref{eq:energyfunctional}) contains the wetting energy $f_w(h)$ and the local free energy of the surfactant-covered free surface $f_s(\Gamma)$. 
Here, $\mathrm{d \mathbf{s}} = \zeta  \mathrm{d \mathbf{x}}$ is the surface element of the curved liquid surface and $\mathrm{d \mathbf{x}}$ is the surface element of the euclidean flat substrate plane as depicted in Fig. \ref{Fig:Biofilm_Surfactant_Sketch}.  
A common choice for the wetting energy is\cite{Pismen2006}
\begin{equation}
 f_w(h) = A \left( - \frac{1}{2h^2} + \frac{h_a^3}{5h^5} \right) \, ,
 \label{eq:wetting_energy}
\end{equation}
which combines destabilizing long-range van-der-Waals and stabilizing short-range interactions. It describes a partially wetting fluid, i.e. a macroscopic drop sitting on a stable adsorption layer of thickness $h_a$.\\
Assuming relatively low densities of surfactant, the contribution of a non-interacting surfactant to the energy of the interface corresponds to an entropic term
\begin{equation}
 f_s(\Gamma) = \gamma + \frac{kT}{a^2} \Gamma \left[ \log(\Gamma)-1 \right] 
 \label{eq:fs}
\end{equation}
that results in the usual linear equation of state. Here, $\gamma$ denotes the surface tension, $kT$ the thermal energy and $a^2$ is the effective area of the surfactant molecules on the interface.
In order to write evolution equations for the film height and the surfactant concentration in the formulation as a gradient dynamics \cite{TAP2012pf, TAP2016prf}, it is necessary to introduce the projection of the area density onto the flat surface of the substrate 
\begin{equation}
\tilde \Gamma(x,y,t)= \zeta \Gamma(x,y,t) \, .
\end{equation}
The free energy functional (\ref{eq:energyfunctional}) can now be used to write evolution equations for  $h(x,y,t)$ and $\tilde \Gamma(x,y,t)$
\begin{align}
 & \partial_t h = \nabla \cdot \left[ Q_{hh} \nabla \frac{\delta F}{\delta h} +  Q_{h\Gamma} \nabla \frac{\delta F}{\delta \tilde \Gamma} \right] \label{eq: gradientdynamics1} \\
 &\partial_t \tilde \Gamma = \nabla \cdot \left[ Q_{\Gamma h} \nabla \frac{\delta F}{\delta h} +  Q_{\Gamma \Gamma} \nabla \frac{\delta F}{\delta \tilde \Gamma} \right]
 \label{eq: gradientdynamics2}
\end{align}
with the positive definite mobility matrix \cite{TAP2012pf, WTG+2015mmnp}
\begin{equation}
 \mathbf{Q} = \begin{pmatrix} Q_{hh} & Q_{\Gamma h} \\   Q_{h \Gamma}
   & Q_{\Gamma \Gamma} \end{pmatrix} = \begin{pmatrix}  \frac{h^3}{3
     \eta} &  \frac{h^2 \Gamma}{2 \eta} \\ \frac{h^2 \Gamma}{2 \eta} &
   \frac{h \Gamma^2}{ \eta} +  D \Gamma \end{pmatrix}\, ,
 \label{eq:Qmatrix}
\end{equation}
where $\eta$ denotes the viscosity of the fluid and $D$ is the diffusivity of surfactant molecules on the interface.
Performing the variations of the free energy functional and considering the thin film limit $\zeta \approx 1$ gives
\begin{align}
 \tfrac{\delta F}{\delta h } &=  \partial_h f_w - \nabla \left[ \omega \nabla h  \right] \\
  \tfrac{\delta F}{\delta \Gamma } &=   \partial_\Gamma f_s  \, .
\end{align}
with $\omega = f_s- \Gamma \partial_\Gamma f_s   = \gamma- \frac{kT}{a^2} \Gamma$. 
Using the common approximation that the change of the surface tension
with surfactant concentration is small as compared to the reference
surface tension $\gamma$ (i.e. $\nabla \left[ \omega \nabla h  \right] \approx \gamma \Delta h$ ), we obtain 
\begin{align}
& \partial_t h = \nabla \cdot \left[ Q_{hh} \nabla [ \partial_h f_w - \gamma\Delta h    ] + Q_{h\Gamma} \nabla (\partial_\Gamma f_s)    \right]  \label{eq:variation1}\\
&  \partial_t \Gamma \approx \partial_t  \tilde \Gamma= \nabla \cdot \left[ Q_{\Gamma h} \nabla [ \partial_h f_w - \gamma\Delta h    ]   +Q_{\Gamma \Gamma} \nabla  (\partial_\Gamma f_s )   \right] \, .
\label{eq:variation2}
\end{align}
The last term of (\ref{eq:variation1}) corresponds to the negative of the Marangoni flux $\vec{j}_M = -\frac{kT}{2 \eta a^2} h^2 \nabla \Gamma$ - a flux in the fluid which is driven by concentration gradients of the surfactant. 
\subsection{Model for surfactant-driven colony spreading}
In order to describe the surfactant-driven spreading of bacterial colonies, the hydrodynamic model equations for passive fluids (\ref{eq:variation1})-(\ref{eq:variation2}) are extended by biological growth and production processes. 
Over time, bacteria will multiply by cell division and possibly
secrete osmolytes. This may result in an influx of water into the
colony caused by the difference of the osmotic pressures in the
film and the underlying moist substrate.\cite{SAW+2012PNASUSA}
We assume that this influx is fast as compared to the growth
processes, which allows us to write biomass production and osmotic influx as one effective growth term $G(h)$.
\footnote{In our previous modelling approach focussing on the osmotic effects and wettability\cite{TJT2016ams, TJL+2017prl} in biofilms, we pointed out that in the limiting case of fast osmotic fluxes, a model which treats water and biomass as two individual fields can be reduced to a simplified model with only one variable for the film height $h$.}
To account for processes such as nutrient and oxygen depletion
\cite{ZSS+2014NJoP, DOP+2013jb} which naturally limit the colony
height, we introduce a critical film height $h^\star$, which
corresponds to the maximal height that can be sustained. It can be related to a local equilibrium of vertical nutrient diffusion and consumption by bacteria.\cite{ZSS+2014NJoP} We assume a simple logistic growth law 
  \begin{align}
G(h) &=   g h \left (1- \frac{h}{h^\star} \right) f_{mod}(h) 
\label{eq:growth}
\end{align}
which is modified locally for very small amounts of biomass by $f_{mod}(h)$
in order to prevent proliferation in the adsorption layer outside of the colony and accounts for the fact that at least one bacterium is needed to start biomass growth.
\footnote{Here, we use $f_{mod}(h) = (1 - \frac{h_u}{h}) [1-\exp(\frac{h_s-h}{h_a})]$,
but other forms of $f_{mod}(h)$ with the same fixed point structure give similar results.
}\\
The second bioactive process that needs to be included in the model
is the production of surfactants by the bacteria. Due to the small
height of the colony as compared to its lateral extension, the surfactant quickly diffuses to the liquid-air interface. 
We thus assume the production rate  of surfactant $P(h,\Gamma)$ to be proportional to the biofilm height, to decrease with increasing surfactant concentration and to cease when the local surfactant concentration reaches a limiting value $\Gamma_\text{max}$:
\begin{equation}
 P(h,\Gamma) = p (\Gamma_\text{max} - \Gamma)  \Theta( \Gamma_\text{max} - \Gamma) \cdot h \Theta(h-h_u) \, .
 \label{eq:surf_production}
\end{equation}
The step-functions $\Theta$ are introduced to ensure that production only 
takes place inside the colony and not in the adsorption layer and that surfactant above the maximal concentration $\Gamma_\mathrm{max}$ is not degraded. \\
We include biomass growth (\ref{eq:growth}) and surfactant production (\ref{eq:surf_production}) as additional non-conserved terms into the evolution equations (\ref{eq:variation1})-(\ref{eq:variation2})
\begin{align}
 \partial_t h =& \nabla \cdot \left[ \tfrac{h^3}{3 \eta}\nabla ( \partial_h f_w - \gamma\Delta h    ) \right]  + \tfrac{k T}{a^2 }\nabla \cdot \left[ \tfrac{h^2}{2 \eta}\nabla \Gamma \right]    + G(h)\label{eq:evolutioneqn1}  \\
  \partial_t \Gamma =& \nabla \cdot \left[\tfrac{h^2 \Gamma}{3 \eta} \nabla ( \partial_h f_w - \gamma\Delta h    )\right] + \tfrac{k T}{a^2 }\nabla \cdot \left[ \tfrac{h \Gamma}{\eta}\nabla \Gamma \right] \notag \\
 &  + \tfrac{k T}{a^2 }D\Delta \Gamma   + P(h,\Gamma) 
\label{eq:evolutioneqn2}
\end{align}
where we have used (\ref{eq:fs}) and (\ref{eq:Qmatrix}).
\begin{figure*}
\centering
\includegraphics[width=0.85\textwidth]{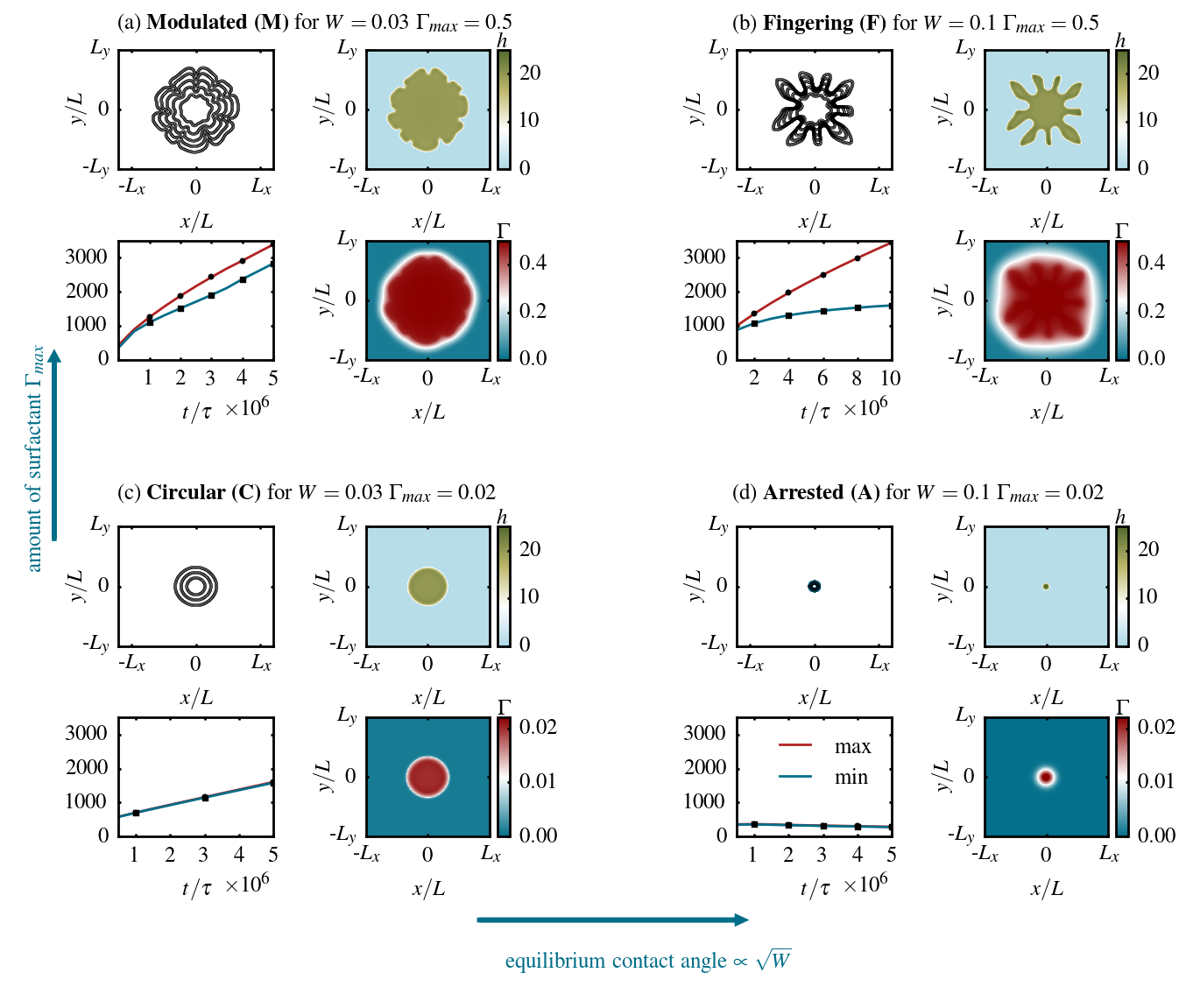}  
 \caption{Panels (a) to (d) show the four different types of spreading that may occur depending on the wettability parameter $W$ and the maximal surfactant concentration $\Gamma_\mathrm{max}$. 
The resp. bottom left plots show the time dependence of the mean values of the maximal and minimal radii of the colony while the resp. top left plots show snapshots of the $h(x,y)=0.5 h^\star$ contour line at the times indicated by filled symbols in the bottom left plots.
The resp. top and bottom right plots show the film height and surfactant distribution, respectively, at the end of the simulation.
Without surfactant, the shape of the bacterial colony is circular. At high wettability (c), the biofilm expands over the substrate with a stable circular front. Under conditions, which do not favour wetting (d), the spreading of the colony is arrested. If the colony produces a significant amount of surfactant $\Gamma_\mathrm{max}$, spreading is promoted by Marangoni fluxes and the circular front becomes unstable. For a high wettability (a), the front continuously advances but is modulated. For a higher value of $W$ (b), the colony forms pronounced fingers which expand over the substrate while the troughs stay arrested.}
  \label{Fig:FourCases}
\end{figure*}
\subsection{Non-dimensional form of the equations}
To obtain a dimensionless form of the model (\ref{eq:evolutioneqn1})-(\ref{eq:evolutioneqn2}) and thereby facilitate the analysis, we introduce the scaling
\begin{equation}
t = \tau \tilde t \qquad x= L \tilde x \qquad y = L\tilde y\qquad h = l \tilde h \qquad f_{w,s} = \kappa \tilde{f}_{w,s}
\end{equation}
where a tilde indicates dimensionless quantities. Time, energy and vertical and horizontal length scales are
\begin{equation}
\tau = \tfrac{L^2 \eta}{\kappa l} \qquad \kappa = \tfrac{kT}{a^2} \qquad l = h_a \qquad  L = \sqrt{\tfrac{\gamma}{\kappa}}l  \, ,
\end{equation}
respectively, and will be estimated quantitatively later in Sec. \ref{sec:timesims_round}.
Inserting the scaling, into the evolution equation results in
the dimensionless biomass growth rate $\tilde g = \tau g$, the dimensionless surfactant production rate $\tilde p = \tau p l$ and the wettability parameter
\begin{equation}
 W = \frac{A a^2}{h_a^2 k_B T} \, ,
\end{equation}
which defines the relative strength of wetting as compared to the entropic influence of the surfactant. It is connected to the equilibrium contact angle $\theta_e$ of passive stationary droplets (without bio-active terms) by $\theta_e \propto \sqrt{W}$ so that larger values of $W$ result in a less wettable substrate and larger contact angles.
If not stated otherwise, we fix the parameters to $\tilde g = 10^{-5}$, $\tilde p = 10^{-6}$, $h^\star = 20.$ and $\tilde D = 0.01$ throughout the analysis and study the effect of the wettability parameter $W$ and the maximal surfactant concentration $\Gamma_\mathrm{max}$ which captures e.g. the difference between a surfactant-producing bacterial strain and a mutant strain deficient in surfactant production.
\section{Results}
\label{sec:results}
In the next section, we present an analysis of the model which focuses on the influence of surfactants and wettability on the spreading dynamics and morphology.
We start by performing time simulations of initially circular colonies at different parameter values $W$ and limiting surfactant concentrations $\Gamma_\mathrm{max}$ as illustrative examples and a graphic description of the effects. These can already be employed to gain a qualitative understanding of the spreading behaviour. In a next step, the spreading regimes are studied for planar fronts by parameter continuation techniques \cite{DO2009}. This results in a more technical description and facilitates, e.g., the analysis of the emerging front instability by a transversal linear stability analysis. The last part of this section contains an illustrative application of the model and exemplarily tests counter-gradients of surfactant as a strategy to arrest the expansion of bacterial colonies. 
\subsection{Influence of wettability and  Marangoni fluxes on the morphology of fronts in radial geometry}
\label{sec:timesims_round}
In a first step, the front dynamics of model (\ref{eq:evolutioneqn1})-(\ref{eq:evolutioneqn2}) is analysed by performing 
two-dimensional numerical time simulations of colony growth. These reveal the influence of wettability and the strength of the Marangoni fluxes on the morphology of the emerging bacterial colonies. We employ a finite element scheme provided by the modular toolbox DUNE-PDELAB\cite{BBD+2008c, BBD+2008ca}.
The simulation domain $\Omega = [-L_x, L_x] \times [-L_y, L_y]$ with $L_x=L_y=5000$ is discretized on a regular mesh using $N_x \times N_y= 512 \times 512$ grid points and linear ansatz and test functions. The time-integration is performed using an implicit second order Runge-Kutta scheme with adaptive time step.  On the boundaries, we apply no-flux conditions for film height and surfactant. The initial condition is given by a small nucleated bacterial colony with surfactant concentration $\Gamma_\mathrm{max}$ on the colony and $0.05\times \Gamma_\mathrm{max}$ on the surrounding substrate. The step functions in the surfactant production term (\ref{eq:surf_production}) are approximated by $\Theta(x) \approx \frac{1}{2}[ 1+ \tanh(100x)]$ .\\
\begin{figure}
\centering
\mbox{
\includegraphics[width=0.5\textwidth]{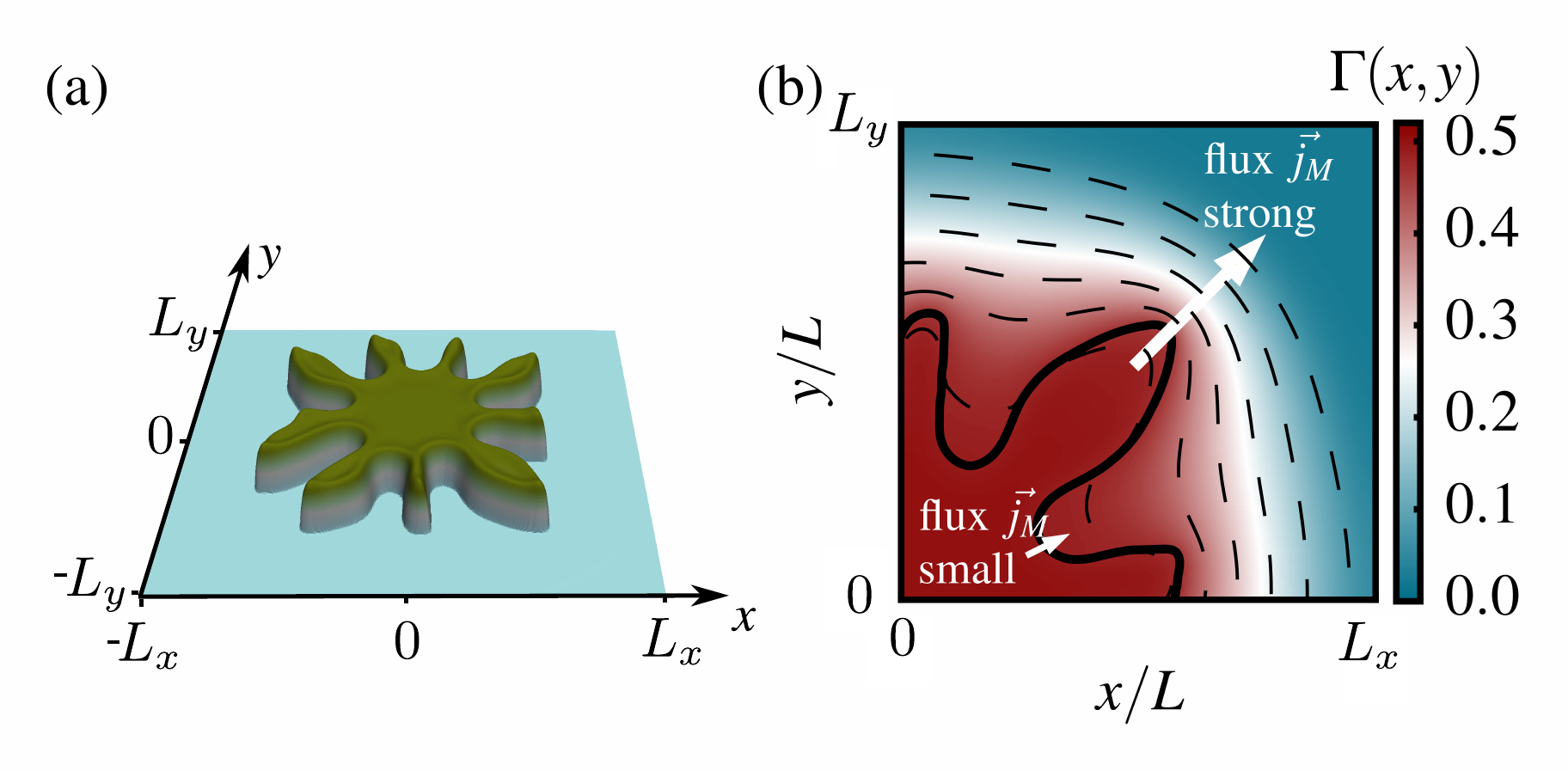}
}
 \caption{Details of the colony which spreads with pronounced fingers for $W=0.1$ and $\Gamma_\mathrm{max}=0.5$. (a) Height profile $h(x,y)$ at $t=10^7 \tau$ reveals the rims which form at the edges of the fingers (cp. Fig. \ref{Fig:FourCases} (b)). (b) Surfactant concentration (colouring, dashed lines represent isolines) on a finger of the colony (solid black line). The gradient in surface tension and thus the Marangoni flux $\vec{j}_M$ is strong at the tips of the fingers, driving them further outwards. In the troughs, the surfactant concentration is at an overall high level so that the Marangoni flux is weak.}
  \label{Fig:DetailsMechanism}
\end{figure}
The time simulations of (\ref{eq:evolutioneqn1})-(\ref{eq:evolutioneqn2}) reveal that - depending on the wettability and the strength of the Marangoni fluxes in the system - four qualitatively different types of spreading behaviour can occur. These are depicted in Fig. \ref{Fig:FourCases} (a) to (d) for four different choices of the parameters $W$ and $\Gamma_\mathrm{max}$. The respective top left plots show the contours of the colonies at equidistant times.
The resp. bottom left plots in Fig. \ref{Fig:FourCases} give the time dependence of the mean values of the maximal and minimal radii of the colony to characterize its shape evolution. The resp. top and bottom right plots show the film height and surfactant distribution profiles at the end of the simulation.\\
We first discuss the spreading behaviour of the system for low surfactant densities (low $\Gamma_\mathrm{max}$). Consistent with the biofilm spreading model without surfactants in \onlinecite{TJL+2017prl}, the system shows a non-equilibrium transition between continuously spreading and arrested colonies depending on the wettability parameter $W$. For small equilibrium contact angles and high wettability (low $W$, Fig.  \ref{Fig:FourCases} (c)), the bacterial colony swells vertically and horizontally until the limiting film height $h^\star$ is reached. Subsequently, it spreads horizontally over the substrate with a constant speed and a \textit{circular} (type C) colony shape. In contrast, at low wettability and thus high contact angle (high $W$, Fig.  \ref{Fig:FourCases} (d)), the spreading of the bacterial colony is \textit{arrested} (type A) and it evolves towards a steady profile of fixed extension and contact angle. \\
In both cases, the production of a significant amount of surfactant (high $\Gamma_\mathrm{max}$) improves the capability of the bacterial colony to expand outwards over the substrate. It results in a higher surfactant concentration $\Gamma(x,y)$ at the centre of the colony than on the surrounding substrate which induces an outwards flow due to the emerging surface tension gradient.  For a continuously spreading colony, these Marangoni flows increase the spreading speed and also cause \textit{modulations} (type M) of the circular colony shape to develop (low $W$, Fig.  \ref{Fig:FourCases} (a)). However, eventually the growth of these undulations slows down and the tips and troughs of the front line translate with a similar velocity over the substrate. This can be clearly seen in the time evolution of the mean values of the maximal and minimal radii of the colony shape. \\
In the case of arrested spreading, the consequences of the surfactant production are even more drastic: In the first place it enables a horizontal expansion of the colony. Furthermore, it gives rise to the formation of pronounced \textit{fingers} (high $W$, type F in Fig.  \ref{Fig:FourCases} (b)). At large times, the finger tips spread outwards with a constant velocity whereas the troughs of the front line stay behind at a fixed position. 
A similar distinction of two types of front instabilities, for which the shape of the evolving front modulations becomes stationary (M) or corresponds to continuously growing fingers (F), has also been made for advancing coating films driven by gravity or shear stress.\cite{ESR2000pf}
The mechanism behind the pronounced fingering mode found here becomes clear when studying the distribution of surfactant on the colony and the surrounding substrate in more detail. Fig. \ref{Fig:DetailsMechanism} (a) shows a height profile of a colony with pronounced fingers at $t=10^7 \tau$. In agreement with the experimental observation \cite{FPB+2012sm}, we find a rim in the height profile at the edges of the colony that is particularly pronounced at the tips of the fingers.  Due to the limiting film height $h^\star$, the centre of the colony is relatively flat. The surfactant concentration is shown in Fig. \ref{Fig:DetailsMechanism} (b) and allows one to understand the formation of the pronounced fingers. In the troughs close to the centre of the colony, the surfactant concentration is overall high and gradients in $\Gamma$ are small. This results in only small Marangoni fluxes which do not suffice to overcome the arrested spreading behaviour. In contrast, at the tips of the fingers, gradients in $\
Gamma$ and 
Marangoni fluxes are strong, driving the finger tips further outwards. If the diffusion of the surfactant is not too high, this gradient in $\Gamma$ is maintained, 
enabling the fingers to continuously spread over the 
substrate. \\
To see if our model predicts a reasonable spreading speed, we estimate the scales for time and length scales by comparing the numerically obtained extensions with experimental measurements and plug in numbers for the remaining constants in the model. The typical colony height of $15 -25\,\mu$m as measured in \onlinecite{FPB+2012sm}, sets the vertical length scale to 
$l = h_a \approx 1\,\mu$m. Together with the viscosity of $\eta \approx 0.1$ Pa s \cite{FPB+2012sm} and surface tension of water $\gamma \approx 70$ mN/m, as well as a typical surfactant length scale $a \approx 3\,$nm, we find the lateral length scale $L\approx 10 \mu$m and the time scale $\tau \approx 0.03$s. With the above scales, our numerically measured dimensionless expansion rate of roughly $5 \times 10^{-4}$ in Fig. \ref{Fig:FourCases} (a) corresponds to a speed of about $10 \mu$m/min, which compares well to the experimentally found value of $5-40 \mu$m/min.\cite{FPB+2012sm} 
\\
To obtain a more complete picture of the front instability, we also study the effect of the remaining parameters $\tilde g$, $\tilde p$, $h^\star$ and $\tilde D$ on the colony shape. We find that a small biomass growth rate $\tilde g$ as well as a small maximal biofilm thickness $h^\star$ promote the instability. Images from the direct time simulations can be found in appendix A1.
\subsection{Profile and velocity of fronts in planar geometry}
\label{SubSec:Continuation}
The time simulations of the two-dimensional system have identified the wettability parameter $W$ and the amount of surfactant $\Gamma_\mathrm{max}$ as two key parameters which influence the spreading of the bacterial colony. To understand the system in more detail, next we investigate planar fronts. In this geometry, it is possible to perform a more systematic analysis of the system using parameter continuation. This technique allows for a direct observation of the influence of $W$ and $\Gamma_\mathrm{max}$ on the front profile and velocity. 
To that end, the evolution equations (\ref{eq:evolutioneqn1})-(\ref{eq:evolutioneqn2}) are transformed into the co-moving coordinate system with a constant velocity $v$
\begin{align}
 \partial_t h =& \nabla \cdot \left[ Q_{hh} \nabla [ \partial_h f_w - \gamma\Delta h    ] + Q_{h\Gamma} \nabla (\partial_\Gamma f_s)    \right] \notag\\
  &+ G(h) + v \partial_x h  \label{eq:comoving1} \\
=& \mathscr{F}_1(\nabla,v)[h, \Gamma] \notag \\
  \partial_t \Gamma =& \nabla \cdot \left[ Q_{\Gamma h} \nabla [ \partial_h f_w - \gamma\Delta h    ]   +Q_{\Gamma \Gamma} \nabla  (\partial_\Gamma f_s )   \right] \notag \\
  &+ P(h,\Gamma) +v \partial_x \Gamma    \label{eq:comoving2} \\
=& \mathscr{F}_2(\nabla,v)[h, \Gamma] \notag \, ,
\end{align}
where we introduced $\mathscr{F}_{1,2}(\nabla, v) $ as a short hand notation for the nonlinear operators defined by the right-hand sides of the evolution equations (\ref{eq:comoving1}) and (\ref{eq:comoving2}). 
In the co-moving frame, planar fronts $(h_0(x),\Gamma_0(x))$ which depend only on one spatial coordinate, $x$, and move with a stationary profile and velocity $v$ correspond to steady solutions 
\begin{align}
 \partial_t h_0(x) &= \mathscr{F}_1(\nabla, v)[h_0(x), \Gamma_0(x)] = 0  \\
  \partial_t \Gamma_0(x) &= \mathscr{F}_2(\nabla,v)[h_0(x), \Gamma_0(x)] =0  \, ,
\end{align}
and can thus be analysed by parameter continuation techniques \cite{DWC+2014ccp, Kuznetsov2013}.
To that end, we use the software package AUTO-07p\cite{DO2009} which has previously been successfully employed for thin-film models, e.g. for dewetting simple and complex liquids \cite{TTL2013PRL}, pattern formation in dip-coating \cite{WZC+2016jopcm} or osmotically spreading biofilms.\cite{TJT2016ams, TJL+2017prl} We impose that far away from the colony, the surfactant concentration is fixed to a small but finite value.\\ 
\begin{figure}
\centering
\includegraphics[width=0.45\textwidth]{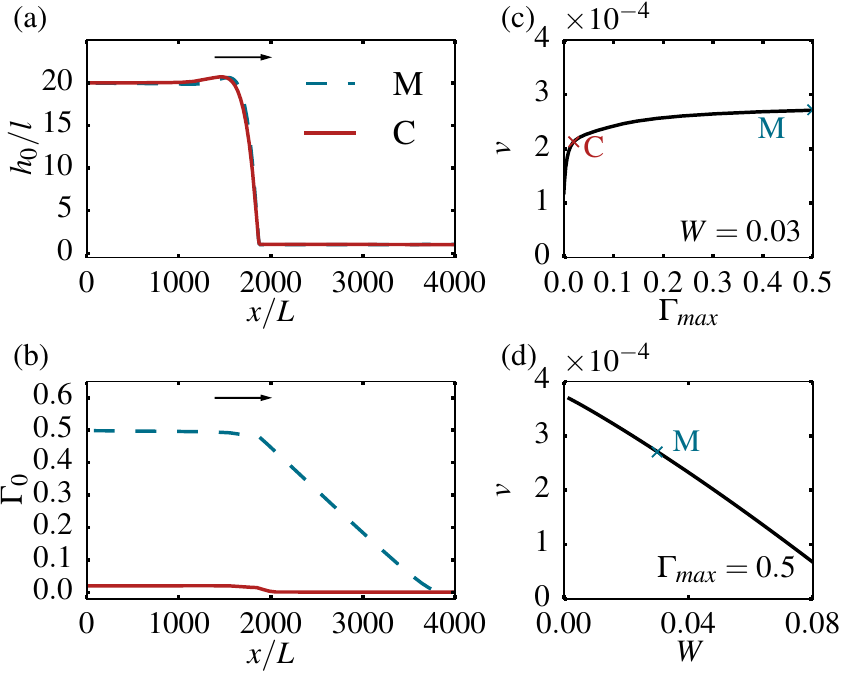}
 \caption{Shape and velocity of planar fronts. (a) and (b): Front profiles for parameter combinations corresponding to the circular (C) colony ($W=0.03$, $\Gamma_\mathrm{max}=0.02$) and the colony with the modulated (M) front shape ($W=0.03$, $\Gamma_\mathrm{max}=0.5$).  The front velocity $v$ strongly depends on the maximal surfactant amount $\Gamma_\mathrm{max}$ (c) as well as the wettability parameter $W$ (d). }
  \label{Fig:Continuation}
\end{figure}
Fig. \ref{Fig:Continuation} (a) and (b) show the front profiles $(h_0(x),\Gamma_0(x))$ for the parameter combinations corresponding to modulated (M) and circular (C) spreading in the radial geometry (Fig \ref{Fig:FourCases}). Behind the spreading front, $h$ and $\Gamma$ reach their saturation values $h^\star$ and $\Gamma_\mathrm{max}$, respectively. The height profile of the front shows a typical capillary rim. The surfactant diffuses in front of the moving colony, resulting in a linear decay. This is a typical profile, as discussed more generally in \onlinecite{WJ2001ijam} for a moving source of surfactant. \\
The velocity of the front is strongly affected by the surfactant concentration $\Gamma_\mathrm{max}$ and the wettability parameter $W$ (see Figs. \ref{Fig:Continuation} (c) and (d)). For a bacterial colony with very small $\Gamma_\mathrm{max}$, e.g. a mutant strain deficient in surfactant production, the front velocity is roughly a factor of two smaller then in a colony with $\Gamma_\mathrm{max}=0.5$. This shows that the Marangoni effect gives a strong contribution to the outward flux that results in the colony expansion. 
In analogy to the transition from continuous to arrested spreading observed in biofilms without Marangoni flows \cite{TJL+2017prl}, the biofilm expansion slows down as the conditions do not favour wetting for large $W$. \\
\\
In the time simulations for radial geometry discussed in section \ref{sec:timesims_round}, we found that the surfactant not only influences the spreading speed of the colonies, but also affects its morphology. We will focus on this aspect in the next section.
\subsection{Transversal linear stability analysis of two-dimensional planar fronts}
\label{SubSec:LTSA}
Now, we analyse the evolution of the morphology of planar fronts in a two-dimensional geometry. To that end, 
we perform a linear transversal stability analysis employing the ansatz
\begin{align}
  h(x,y,t) &=  h_0(x) +  \epsilon h_{1}(x) \exp(i k_y y + \sigma t) \label{eq:LTSA1}\\
  \Gamma(x,y,t) &= \Gamma_0(x) + \epsilon \Gamma_{1}(x) \exp(i k_y y + \sigma t) \label{eq:LTSA2} 
 \end{align}
 with $\epsilon \ll 1$. 
 This corresponds to fronts consisting of a $y$-invariant base state given by the stationary fronts $(h_0(x),\Gamma_0(x))$ plus a small perturbation with $x$-dependence $(h_1(x), \Gamma_1(x))$ which is modulated in the $y$-direction with a wavenumber $k_y$ and grows or decays exponentially in time with the rate $\sigma$.
Inserting this ansatz into the evolution equations (\ref{eq:comoving1})-(\ref{eq:comoving2}) one obtains to $O(\epsilon)$ the linear eigenvalue problem 
\begin{align}
\sigma h_{1}(x)  &= \mathscr{F}_{1h}' |_{h_0(x), \Gamma_0(x)} h_{1}(x) + \mathscr{F}_{1\Gamma}' |_{h_0(x), \Gamma_0(x)} \Gamma_{1}(x)  \label{eq:LTSA_ev1}  \\
\sigma \Gamma_{1}(x)&= \mathscr{F}_{2h}' |_{h_0(x), \Gamma_0(x)} h_{1}(x)+ \mathscr{F}_{2\Gamma}' |_{h_0(x), \Gamma_0(x)} \Gamma_{1}(x) \label{eq:LTSA_ev2}  \, ,
\end{align}
for eigenvalues $\sigma$ and eigenfunctions $(h_1(x), \Gamma_1(x))$, where $\mathscr{F}'_{i,h}$ and $\mathscr{F}'_{i,\Gamma}$ are operators denoting the Fr\'echet-derivatives of the non-linear operator  $\mathscr{F}_i$ with respect to $h$ and $\Gamma$, respectively. Statements about the linear stability of the front $(h_0(x),\Gamma_0(x))$ can now be made determining the largest eigenvalue $\sigma$ which tells if the perturbation grows (for $\sigma >0$) or decays (for $\sigma <0$) in time. 
The linear eigenvalue problem (\ref{eq:LTSA_ev1})-(\ref{eq:LTSA_ev2}) is again solved using continuation techniques. The set of equations for the steady front profiles $h_0(x)$ and $\Gamma_0(x)$ employed in section \ref{SubSec:Continuation} is supplemented by a set of equations for the eigenfunctions that fulfill the same boundary conditions as the base state. This approach, in which transversal wave number and eigenvalue are treated as parameters in an pseudo-arclength continuation, is presented in tutorial form in Ref. \onlinecite{cenosTutorialLindrop}.\\
\begin{figure}
\centering
\includegraphics[width=0.45\textwidth]{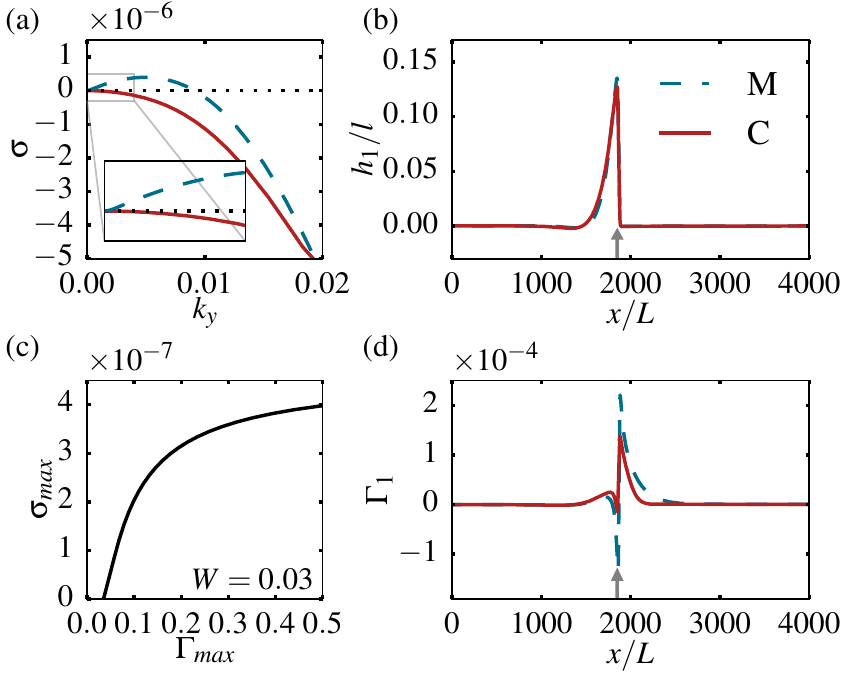}
 \caption{Linear transversal stability analysis for planar fronts in the co-moving frame for $W=0.03$.  The dispersion relation shown in (a) monotonically decreases for $\Gamma_\mathrm{max} = 0.02$ (red) but has a maximum $\sigma_\mathrm{max}>0$ for $\Gamma_\mathrm{max}=0.5$ (blue) indicating a front that is unstable with respect to transversal perturbations. (b) and (d) show the eigenfunctions for the fronts profiles in Fig. \ref{Fig:Continuation} corresponding to the transversal wave number $k_y$ with the largest eigenvalue. The grey arrow indicates the front position of the corresponding $h$ profiles. Following the maximum of the dispersion relation $\sigma_\mathrm{max}$ depending on $\Gamma_\mathrm{max}$ in (c) shows that for $W=0.03§,$ the front profile is unstable for $\Gamma_\mathrm{max} > 0.0268$. }
  \label{Fig:LTSA}
 \end{figure} 
 We again investigate the front profiles for two parameter sets which correspond to the modulated (M) and circular (C) spreading in the radial geometry (Fig. \ref{Fig:FourCases}). Recall that the respective base states $(h_0(x),\Gamma_0(x))$ are displayed in Fig. \ref{Fig:Continuation}. In order to determine the transversal stability of these fronts, one needs to analyse the corresponding dispersion relations $\sigma(k_y)$ which are shown in Fig. \ref{Fig:LTSA} (a).
For the parameter set (C) with only a small concentration of surfactant $\Gamma_\mathrm{max}=0.02$, the dispersion relation decays monotonically (red solid line in Fig. \ref{Fig:LTSA} (a)). The largest eigenvalue is  $\sigma_\mathrm{max}=0$ at $k_y=0$ and the front is thus transversally stable. The eigenfunction $(h_1(x),\Gamma_1(x))$ corresponding to the largest eigenvalue (red solid lines in Fig. \ref{Fig:LTSA} (b),(d)) is the neutrally stable (Goldstone) mode representing the translational symmetry of the equations. As expected, it is identical to the spatial derivative of the front profiles (data not shown). \\
For the other parameter set (M), which corresponds to the situation that a significant amount of surfactant $\Gamma_\mathrm{max}=0.5$ is present in the system, the largest eigenvalue is positive at finite wavenumber  ($\sigma_\mathrm{max}=3.98\times 10^{-7} $ at $k_y =4.89\times 10^{-3}$) and the front is thus transversally unstable (blue dashed line in Fig. \ref{Fig:LTSA} (a)).
These values roughly agree with linear results extracted from the fully nonlinear time simulation in a planar (data not shown). 
The eigenfunctions corresponding to the largest eigenvalue (blue dashed lines in \ref{Fig:LTSA} (b),(d)) are strongly localized in the front region (compare to Fig. \ref{Fig:Continuation}). 
To find the surfactant concentration at which the transition from transversally stable to unstable fronts takes place, we follow the maximum of the dispersion relation $\sigma_\mathrm{max}$ while varying $\Gamma_\mathrm{max}$ (Fig. \ref{Fig:LTSA} (c)). At $W=0.03$, we find that $\sigma_\mathrm{max}$ is positive and the front profile thus transversally unstable for $\Gamma_\mathrm{max}>0.0268$. 
\subsection{Comparison to surfactant-driven spreading of 'passive' thin films}
The front profiles observed in our model show some of the main characteristics of the solutions observed for surfactant-driven spreading of passive thin liquid films as e.g. the capillary rim near the edge of the front and the linear decay of the surfactant concentration in front of the drop. 
In contrast to other modelling approaches \cite{JN2006jofm, CM2006pof, WCM2004pof}, we incorporate a wetting energy corresponding to a partially wetting fluid, resulting in a stable adsorption layer of height $h_a$ independent of the surfactant concentration $\Gamma$. Therefore, we do not observe the typical fluid step or a thinned region in front of the advancing colony which is often described as the origin of the front instability observed for surfactant-driven spreading of 'passive' drops on horizontal substrates.\cite{MC2009sm}  Instead, our model for surfactant-driven colony spreading shows similarity to a surfactant covered drop sliding down an inclined substrate.\cite{EMC2004joem, EMC2005pdnp, GN2015jofm} 
In this set-up, the spreading of the drop is - in addition to the Marangoni fluxes - driven by gravity which acts as a body force on the fluid. In our model, the driving is presented by the non-conserved biomass growth term.
The fingering instability and the eigenfunctions of the unstable mode observed in our model strongly resemble the transversal perturbations found in a constant-flux configuration in \onlinecite{EMC2005pdnp} which are also located at the front edge rather than in the region ahead of it. 
\\
\subsection{Phase-diagram for planar fronts}
\label{SubSec:Phasediagram}
\begin{figure}
\centering
\includegraphics[width=0.45\textwidth]{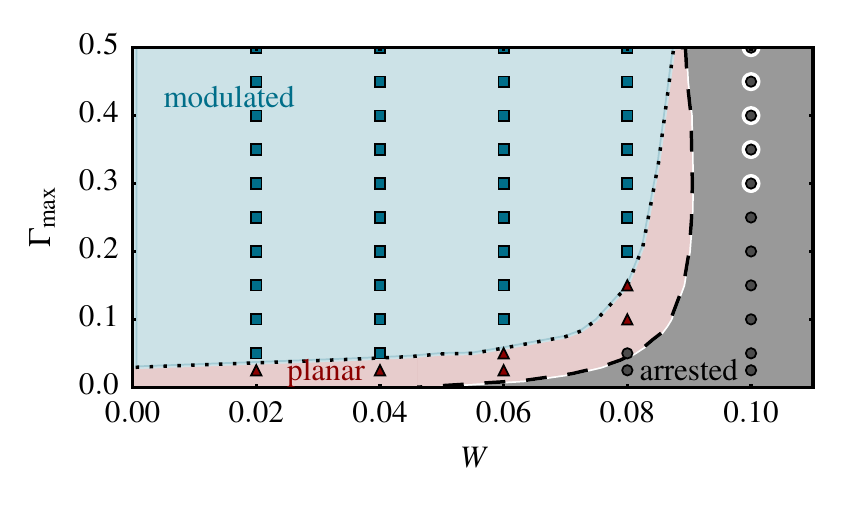}
\caption{Phase-diagram for the spreading of two-dimensional planar fronts. To the left (right) of the dashed line, the fronts of the height profile $h$ are moving (not moving). The linear stability analysis predicts modulated moving fronts to the left of the dotted line (the largest eigenvalue $\sigma_\mathrm{max}$ is $> 10^{-8}$). In numerical time simulations initiated with a noisy planar front, three types of spreading occur: transversally stable planar fronts (red triangles), modulated fronts (blue squares) and arrested fronts which can not expand over the substrate (grey dots). The grey dots with white circles mark parameter sets for which a finite perturbation in the initial condition leads to spreading of a pronounced finger.\label{Fig:Phasediagram}}
\end{figure}

We complete our analysis of planar fronts by combining our results from the transversal linear stability analysis with time simulations. This allows us to determine a phase-diagram which distinguishes between different spreading modes depending on the wettability parameter $W$ and the surfactant concentration $\Gamma_\mathrm{max}$. 
The time simulations are performed on a domain $\Omega = [-L_x, L_x] \times [-L_y, L_y]$ with $L_x=1500$ and $L_y=3000$ discretized on an equidistant $N_x \times N_y= 256 \times 512$ mesh with the same integration method and boundary conditions as applied in section \ref{sec:timesims_round}. The initial condition consists of a noisy planar front given by the corresponding stationary front profile $(h_0(x), \Gamma_0(x))$ for each parameter set. The simulation time is $T_\mathrm{end}/\tau=10^7 $. For initially planar fronts, we find three different spreading regimes as shown in Fig. \ref{Fig:Phasediagram}: Arrested planar fronts, which do not advance (grey dots), moving planar fronts (red triangles) and moving modulated fronts (blue squares) for which the transversal perturbations $\Delta h$ grow in time (and thus $\frac{\Delta h (t=T_\mathrm{end})}{\Delta h (t=0)} >1$). 
We find the same tendencies as observed in the circular geometry in section \ref{sec:timesims_round}. At low surfactant concentration, the front spreads without a transversal instability for a small contact angle (low $W$) but is arrested for high $W$. An increased surfactant concentration 
leads to a modulated front. We compare this findings with the predictions from the transversal linear stability analysis. We identify the region in which the ansatz (\ref{eq:LTSA1})-(\ref{eq:LTSA2}) of moving fronts is valid (stationary $h$ and $\Gamma$ in the co-moving frame) is valid.
In the grey region to the right of the dashed line in Fig. \ref{Fig:Phasediagram}, this condition breaks down and we do not expect a stationary moving front. This is in accordance with the occurrence of the arrested mode in the time simulations. Note that in this situation, the produced surfactant still spreads outwards and the arrested growth mode does therefore not correspond to a stationary front with $v=0$ for both fields $h$ and $\Gamma$. The transversal linear stability makes a prediction about the strength of the transversal instability via the largest eigenvalue $\sigma_\mathrm{max}$. To the left of the dotted line in Fig. \ref{Fig:Phasediagram}, the eigenvalue is larger than $\sigma_\mathrm{max}=10^{-8}$ and the 
modulation of the front should be observable within our simulation time $T_{end} /\tau= 5 \cdot 10^6 .. 10^7 $. This is in good agreement with the time simulations.

Interestingly, the fingering mode (F) does not occur for time simulations initiated with planar fronts that are only slightly perturbed. In general, the transversal instability appears to be much weaker then observed in the radial geometry. This can be attributed to a dilution effect of the surfactant: In the radial geometry, the produced surfactant is diluted more strongly when it spreads outwards from the colony and the surfactant profile decays faster with the distance to the front. This results in stronger gradients in surfactant concentration which drive the transversal instability. 
To test if the fingering mode only exists for circular colonies, we perform the time simulations with an initial condition consisting of a planar front with a finite-size perturbation in the form of a small finger. We find that at large surfactant concentrations, the arrested spreading mode can be overcome (grey dots with white circle in Fig. \ref{Fig:Phasediagram}): the initial finger continuously grows while the rest of the front stays behind similar to the radial geometry. In conclusion, the instability and especially the fingering mode 
are generically occurring in the planar and the radial geometry, however, the onset and strength critically depend on colony shape.
\subsection{Preventing the growth of bacterial colonies by a counter-gradient of surfactant}
\begin{figure}
\centering
\includegraphics[width=0.45\textwidth]{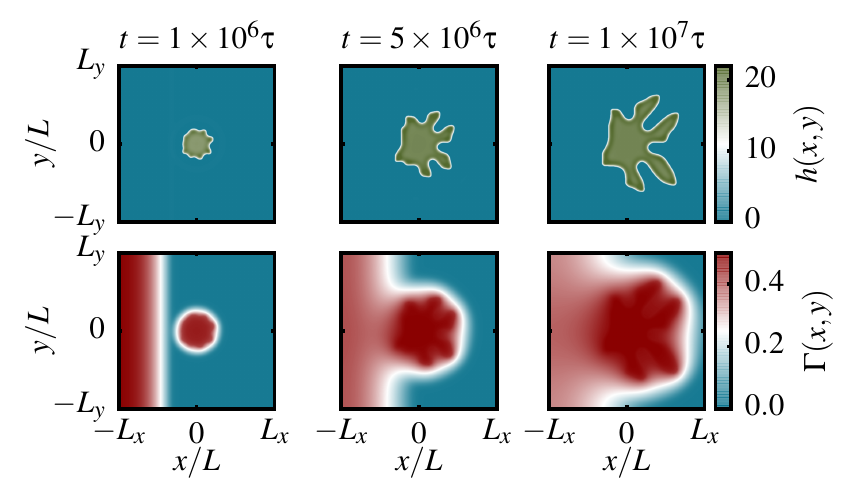}
 \caption{Inhibition of bacterial expansion by a counter-gradient of surfactant. Initially, surfactant is deposited on the left border of the computational domain, while a small drop of a bacterial colony is initiated in the centre. The counter-gradient of external surfactant inhibits fingering towards the left border. Parameters are $W=0.1$ and $\Gamma_\mathrm{max}=0.5$.}
  \label{Fig:Inhibition}
 \end{figure}

After analysing the model mathematically in sections \ref{SubSec:Continuation} to \ref{SubSec:Phasediagram}, we illustrate the consequences of the spreading mechanism at an example. 
One strategy that has been suggested to arrest the expansion of a bacterial colony are counter-gradients of surfactants.
Indeed, experiments\cite{FPB+2012sm,CSO2005job} show that spreading of a \textit{P. aeruginosa} colony can be inhibited if exogenously added bio-surfactant is present on the agar substrate in a circular pattern around the colony with a concentration comparable to the \textit{in vivo} one.
To test this strategy in our model, we perform a time simulation with $W=0.1$ and $\Gamma_\mathrm{max}=0.5$ starting from an initial condition given by a surfactant-laden colony at the centre of the simulation domain and additional surfactant to the left of the colony. The time simulation (see Fig. \ref{Fig:Inhibition}) shows that the growth of the colony towards the left hand side slows down as soon as the colony 'senses' the additional surfactant and eventually its growth is arrested. At the other side, the colony performs the expected finger-like growth. This effect can also be expected to occur when two surfactant-producing colonies approach each other, as observed experimentally\cite{TRL+2007em}. 
\section{Conclusion and Outlook}
We have developed and studied a simple model for surfactant-driven biofilm spreading which demonstrates that wettability and Marangoni fluxes have a strong effect on the expansion behaviour and morphology of bacterial colonies.  The model we have presented is based on a hydrodynamic approach including wetting forces supplemented by bioactive terms and thus allows us to study the interplay between biological growth processes and passive surface forces. 
We find four different types of spreading, ranging from arrested spreading over circular spreading and undulated spreading fronts to the formation of pronounced fingers. The obtained results show that the production of bio-surfactants can enable a bacterial colony to spread over the substrate under conditions, which are otherwise unfavourable to a horizontal expansion; This is because the resulting Marangoni fluxes can significantly contribute to the spreading velocity.
Our results are in qualitative agreement with the experimental findings \cite{FPB+2012sm, CSO2005job, TRL+2007em} which show that surfactant-producing \textit{Pseusomona aeruginosa} colonies spread outwards and form pronounced fingers whereas a wild type deficient in surfactant production can not expand and is arrested in a small circular shape. This corresponds to the transition from the fingering mode to the arrested spreading mode that our model predicts.\\

Note that, here, capillarity and wettability on the one hand and the gradients in surface tension on the other hand have been treated separately in order to discuss their respective effects. In an experiment with bacterial colonies on agar plates, the wettability also depends on the concentration of surfactants because they alter the surface tension and thus the Hamaker constant (entering the parameter $W$)\cite{TST+2018apa}. Therefore, the difference between a bacterial strain deficient in surfactant production and a surfactant producing strain implies that the latter has a lower parameter $W$ and a higher surfactant concentration $\Gamma_\mathrm{max}$ (see discussion in \onlinecite{TJL+2017prl}).\\
In this work, we have followed a simple two-field approach, treating the bacterial colony as a complex fluid covered by surfactants. To capture situations, in which variations of the colony composition are not negligible, e.g. because of similar time scales for biomass growth and osmotic processes, the model can be extended to a three-field model. There, the water concentration enters as a separate field as described in \onlinecite{TJL+2017prl}. 
In addition, the extension of the model to soluble surfactants with a bulk concentration is straight forward, following the model for passive fluids presented in \onlinecite{TAP2016prf}.
 Our modelling approach neglects complex features, such as vertical gradients or cell differentiation. Experiments \cite{FPB+2012sm} in which the
  \textit{rhamnolipid} production in \textit{Pseudomona aeruginosa}
  colonies is highlighted by autofluorescence indicate that there are
  only small spatio-temporal variations in surfactant production
  throughout the colony but in general, cell differentiation is an important phenomenon in bacterial colonies and biofilms\cite{VAL+2008gd}.
In future extensions of the model, one may also incorporate the quorum sensing role of the bio-surfactants which allows for a basic form of communication between individual cells. However, as our model focuses on the physical effects of bio-surfactants, it is well suited to show that these suffice to induce the striking fingering colony shapes which are observed experimentally.
\section*{Conflict of interest}
There are no conflicts to declare.

\section*{Acknowledgement}
We thank the DAAD, Studienstiftung des deutschen Volkes, 
Campus France (PHC PROCOPE grant~35488SJ) and the CNRS (grant PICS07343) for financial
support. LIPhy is part of LabEx~Tec~21 (Invest.\ l'Avenir,
grant~ANR-11-LABX-0030).


%

\section{Appendix}
\begin{figure*}
\centering
 \includegraphics[width=0.7\textwidth]{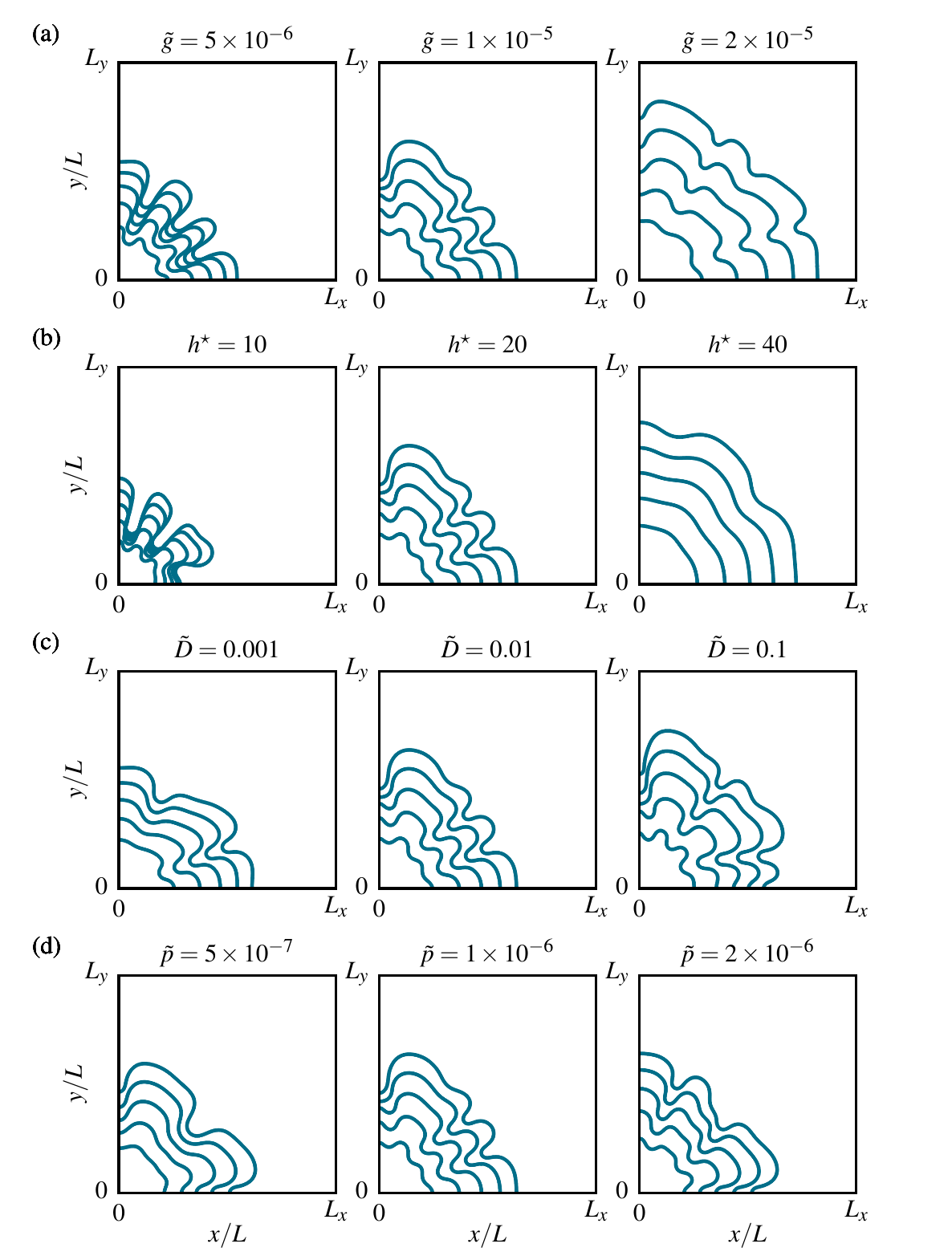}        
 \caption{Parameter study to test the influence of the remaining parameters $\tilde g$, $h^\star$, $\tilde D$ and $\tilde p$ on the colony shape. The contour line $h(x,y)=0.5 h^\star$ is shown at equidistant times with $\Delta t = 10^6$. In each run, one parameter is varied as compared to a reference parameter set (middle column) with $W=0.05$, $\Gamma_\mathrm{max}=0.5$, $\tilde g = 10^{-5}$, $\tilde p = 10^{-6}$, $h^\star = 20.$ and $\tilde D = 0.01$.  }
   \label{Fig:Appendix}
\end{figure*}

For completeness, we briefly investigate the influence of the remaining dimensionless parameters and test the robustness of the observed phenomena by performing additional numerical time simulations. A simulation with the parameters $W=0.05$,  $\Gamma_\mathrm{max}=0.5$, $\tilde g = 10^{-5}$, $\tilde p = 10^{-6}$, $h^\star = 20.$ and $\tilde D = 0.01$ is used as a reference and each parameter is varied individually. Fig. \ref{Fig:Appendix} shows time simulations which are initiated with a quarter of a small bacterial colony with surfactant concentration $\Gamma_\mathrm{max}$ on a domain $\Omega = [0, 5000] \times [0, 5000]$ discretized on an equidistant mesh of $N_x \times N_y= 256 \times 256$ grid points.
A larger biomass production rate $\tilde g$ reduces the formation of fingers as the hydrodynamic time-scale which is relevant for the Marangoni fluxes that drive the instability is no longer fast enough (see Fig.\ref{Fig:Appendix} (a)). An increase of the limiting height $h^\star$ also results in a weakening of the instability (see Fig.\ref{Fig:Appendix} (b)). A change of the surfactant diffusion $\tilde D$ or the production rate $\tilde p$ does not change the morphology of the colonies drastically (see Fig.\ref{Fig:Appendix} (b) and (c), respectively).


\begin{thebibliography}{60}%
\makeatletter
\providecommand \@ifxundefined [1]{%
 \@ifx{#1\undefined}
}%
\providecommand \@ifnum [1]{%
 \ifnum #1\expandafter \@firstoftwo
 \else \expandafter \@secondoftwo
 \fi
}%
\providecommand \@ifx [1]{%
 \ifx #1\expandafter \@firstoftwo
 \else \expandafter \@secondoftwo
 \fi
}%
\providecommand \natexlab [1]{#1}%
\providecommand \enquote  [1]{``#1''}%
\providecommand \bibnamefont  [1]{#1}%
\providecommand \bibfnamefont [1]{#1}%
\providecommand \citenamefont [1]{#1}%
\providecommand \href@noop [0]{\@secondoftwo}%
\providecommand \href [0]{\begingroup \@sanitize@url \@href}%
\providecommand \@href[1]{\@@startlink{#1}\@@href}%
\providecommand \@@href[1]{\endgroup#1\@@endlink}%
\providecommand \@sanitize@url [0]{\catcode `\\12\catcode `\$12\catcode
  `\&12\catcode `\#12\catcode `\^12\catcode `\_12\catcode `\%12\relax}%
\providecommand \@@startlink[1]{}%
\providecommand \@@endlink[0]{}%
\providecommand \url  [0]{\begingroup\@sanitize@url \@url }%
\providecommand \@url [1]{\endgroup\@href {#1}{\urlprefix }}%
\providecommand \urlprefix  [0]{URL }%
\providecommand \Eprint [0]{\href }%
\providecommand \doibase [0]{http://dx.doi.org/}%
\providecommand \selectlanguage [0]{\@gobble}%
\providecommand \bibinfo  [0]{\@secondoftwo}%
\providecommand \bibfield  [0]{\@secondoftwo}%
\providecommand \translation [1]{[#1]}%
\providecommand \BibitemOpen [0]{}%
\providecommand \bibitemStop [0]{}%
\providecommand \bibitemNoStop [0]{.\EOS\space}%
\providecommand \EOS [0]{\spacefactor3000\relax}%
\providecommand \BibitemShut  [1]{\csname bibitem#1\endcsname}%
\let\auto@bib@innerbib\@empty
\bibitem [{\citenamefont {Donlan}(2002)}]{Donl2002eid}%
  \BibitemOpen
  \bibfield  {author} {\bibinfo {author} {\bibfnamefont {R.~M.}\ \bibnamefont
  {Donlan}},\ }\href {\doibase DOI: 10.3201/eid0809.020063} {\bibfield
  {journal} {\bibinfo  {journal} {Emerg. Infect. Dis.}\ }\textbf {\bibinfo
  {volume} {8}},\ \bibinfo {pages} {881} (\bibinfo {year} {2002})}\BibitemShut
  {NoStop}%
\bibitem [{\citenamefont {Stoodley}\ \emph {et~al.}(2002)\citenamefont
  {Stoodley}, \citenamefont {Sauer}, \citenamefont {Davies},\ and\
  \citenamefont {Costerton}}]{SSD+2002ARoM}%
  \BibitemOpen
  \bibfield  {author} {\bibinfo {author} {\bibfnamefont {P.}~\bibnamefont
  {Stoodley}}, \bibinfo {author} {\bibfnamefont {K.}~\bibnamefont {Sauer}},
  \bibinfo {author} {\bibfnamefont {D.~G.}\ \bibnamefont {Davies}}, \ and\
  \bibinfo {author} {\bibfnamefont {J.~W.}\ \bibnamefont {Costerton}},\ }\href
  {\doibase 10.1146/annurev.micro.56.012302.160705} {\bibfield  {journal}
  {\bibinfo  {journal} {Annu. Rev. Microbiol.}\ }\textbf {\bibinfo {volume}
  {56}},\ \bibinfo {pages} {187} (\bibinfo {year} {2002})}\BibitemShut
  {NoStop}%
\bibitem [{\citenamefont {Yang}\ \emph {et~al.}(2017)\citenamefont {Yang},
  \citenamefont {Tang}, \citenamefont {Si},\ and\ \citenamefont
  {Tang}}]{YTS+2017bj}%
  \BibitemOpen
  \bibfield  {author} {\bibinfo {author} {\bibfnamefont {A.}~\bibnamefont
  {Yang}}, \bibinfo {author} {\bibfnamefont {W.~S.}\ \bibnamefont {Tang}},
  \bibinfo {author} {\bibfnamefont {T.}~\bibnamefont {Si}}, \ and\ \bibinfo
  {author} {\bibfnamefont {J.~X.}\ \bibnamefont {Tang}},\ }\href@noop {}
  {\bibfield  {journal} {\bibinfo  {journal} {Biophys. J.}\ }\textbf {\bibinfo
  {volume} {112}},\ \bibinfo {pages} {1462} (\bibinfo {year}
  {2017})}\BibitemShut {NoStop}%
\bibitem [{\citenamefont {Seminara}\ \emph {et~al.}(2012)\citenamefont
  {Seminara}, \citenamefont {Angelini}, \citenamefont {Wilking}, \citenamefont
  {Vlamakis}, \citenamefont {Ebrahim}, \citenamefont {Kolter}, \citenamefont
  {Weitz},\ and\ \citenamefont {Brenner}}]{SAW+2012PNASUSA}%
  \BibitemOpen
  \bibfield  {author} {\bibinfo {author} {\bibfnamefont {A.}~\bibnamefont
  {Seminara}}, \bibinfo {author} {\bibfnamefont {T.}~\bibnamefont {Angelini}},
  \bibinfo {author} {\bibfnamefont {J.}~\bibnamefont {Wilking}}, \bibinfo
  {author} {\bibfnamefont {H.}~\bibnamefont {Vlamakis}}, \bibinfo {author}
  {\bibfnamefont {S.}~\bibnamefont {Ebrahim}}, \bibinfo {author} {\bibfnamefont
  {R.}~\bibnamefont {Kolter}}, \bibinfo {author} {\bibfnamefont
  {D.}~\bibnamefont {Weitz}}, \ and\ \bibinfo {author} {\bibfnamefont
  {M.}~\bibnamefont {Brenner}},\ }\href {\doibase 10.1073/pnas.1109261108}
  {\bibfield  {journal} {\bibinfo  {journal} {Proc. Natl. Acad. Sci. U. S. A.}\
  }\textbf {\bibinfo {volume} {109}},\ \bibinfo {pages} {1116} (\bibinfo {year}
  {2012})}\BibitemShut {NoStop}%
\bibitem [{\citenamefont {Trinschek}\ \emph {et~al.}(2016)\citenamefont
  {Trinschek}, \citenamefont {John},\ and\ \citenamefont
  {Thiele}}]{TJT2016ams}%
  \BibitemOpen
  \bibfield  {author} {\bibinfo {author} {\bibfnamefont {S.}~\bibnamefont
  {Trinschek}}, \bibinfo {author} {\bibfnamefont {K.}~\bibnamefont {John}}, \
  and\ \bibinfo {author} {\bibfnamefont {U.}~\bibnamefont {Thiele}},\ }\href
  {\doibase http://dx.doi.org/10.3934/matersci.2016.3.1138} {\bibfield
  {journal} {\bibinfo  {journal} {AIMS Materials Science}\ }\textbf {\bibinfo
  {volume} {3}},\ \bibinfo {pages} {1138} (\bibinfo {year} {2016})}\BibitemShut
  {NoStop}%
\bibitem [{\citenamefont {Dilanji}\ \emph {et~al.}(2014)\citenamefont
  {Dilanji}, \citenamefont {Teplitski},\ and\ \citenamefont
  {Hagen}}]{DTH2014prsb}%
  \BibitemOpen
  \bibfield  {author} {\bibinfo {author} {\bibfnamefont {G.~E.}\ \bibnamefont
  {Dilanji}}, \bibinfo {author} {\bibfnamefont {M.}~\bibnamefont {Teplitski}},
  \ and\ \bibinfo {author} {\bibfnamefont {S.~J.}\ \bibnamefont {Hagen}},\
  }\href@noop {} {\bibfield  {journal} {\bibinfo  {journal} {Proc. R. Soc. B}\
  }\textbf {\bibinfo {volume} {281}},\ \bibinfo {pages} {1784} (\bibinfo {year}
  {2014})}\BibitemShut {NoStop}%
\bibitem [{\citenamefont {Yan}\ \emph {et~al.}(2017)\citenamefont {Yan},
  \citenamefont {Nadell}, \citenamefont {Stone}, \citenamefont {Wingreen},\
  and\ \citenamefont {Bassler}}]{YNS+2017nc}%
  \BibitemOpen
  \bibfield  {author} {\bibinfo {author} {\bibfnamefont {J.}~\bibnamefont
  {Yan}}, \bibinfo {author} {\bibfnamefont {C.~D.}\ \bibnamefont {Nadell}},
  \bibinfo {author} {\bibfnamefont {H.~A.}\ \bibnamefont {Stone}}, \bibinfo
  {author} {\bibfnamefont {N.~S.}\ \bibnamefont {Wingreen}}, \ and\ \bibinfo
  {author} {\bibfnamefont {B.~L.}\ \bibnamefont {Bassler}},\ }\href@noop {}
  {\bibfield  {journal} {\bibinfo  {journal} {Nature Comm.}\ }\textbf {\bibinfo
  {volume} {8}},\ \bibinfo {pages} {327} (\bibinfo {year} {2017})}\BibitemShut
  {NoStop}%
\bibitem [{\citenamefont {Ron}\ and\ \citenamefont
  {Rosenberg}(2001)}]{RR2001Em}%
  \BibitemOpen
  \bibfield  {author} {\bibinfo {author} {\bibfnamefont {E.~Z.}\ \bibnamefont
  {Ron}}\ and\ \bibinfo {author} {\bibfnamefont {E.}~\bibnamefont
  {Rosenberg}},\ }\href@noop {} {\bibfield  {journal} {\bibinfo  {journal}
  {Environ. Microbiol.}\ }\textbf {\bibinfo {volume} {3}},\ \bibinfo {pages}
  {229} (\bibinfo {year} {2001})}\BibitemShut {NoStop}%
\bibitem [{\citenamefont {Raaijmakers}\ \emph {et~al.}(2010)\citenamefont
  {Raaijmakers}, \citenamefont {De~Bruijn}, \citenamefont {Nybroe},\ and\
  \citenamefont {Ongena}}]{RDN+2010Fmr}%
  \BibitemOpen
  \bibfield  {author} {\bibinfo {author} {\bibfnamefont {J.~M.}\ \bibnamefont
  {Raaijmakers}}, \bibinfo {author} {\bibfnamefont {I.}~\bibnamefont
  {De~Bruijn}}, \bibinfo {author} {\bibfnamefont {O.}~\bibnamefont {Nybroe}}, \
  and\ \bibinfo {author} {\bibfnamefont {M.}~\bibnamefont {Ongena}},\
  }\href@noop {} {\bibfield  {journal} {\bibinfo  {journal} {FEMS Microbiol.
  Rev.}\ }\textbf {\bibinfo {volume} {34}},\ \bibinfo {pages} {1037} (\bibinfo
  {year} {2010})}\BibitemShut {NoStop}%
\bibitem [{\citenamefont {Ke}\ \emph {et~al.}(2015)\citenamefont {Ke},
  \citenamefont {Hsueh}, \citenamefont {Cheng}, \citenamefont {Wu},\ and\
  \citenamefont {Liu}}]{KHC+2015fm}%
  \BibitemOpen
  \bibfield  {author} {\bibinfo {author} {\bibfnamefont {W.-J.}\ \bibnamefont
  {Ke}}, \bibinfo {author} {\bibfnamefont {Y.-H.}\ \bibnamefont {Hsueh}},
  \bibinfo {author} {\bibfnamefont {Y.-C.}\ \bibnamefont {Cheng}}, \bibinfo
  {author} {\bibfnamefont {C.-C.}\ \bibnamefont {Wu}}, \ and\ \bibinfo {author}
  {\bibfnamefont {S.-T.}\ \bibnamefont {Liu}},\ }\href {\doibase
  10.3389/fmicb.2015.01017} {\bibfield  {journal} {\bibinfo  {journal} {Front
  Microbiol}\ }\textbf {\bibinfo {volume} {6}},\ \bibinfo {pages} {1017}
  (\bibinfo {year} {2015})}\BibitemShut {NoStop}%
\bibitem [{\citenamefont {Lecl{\`e}re}\ \emph {et~al.}(2006)\citenamefont
  {Lecl{\`e}re}, \citenamefont {Marti}, \citenamefont {B{\'e}chet},
  \citenamefont {Fickers},\ and\ \citenamefont {Jacques}}]{LMB+2006AoM}%
  \BibitemOpen
  \bibfield  {author} {\bibinfo {author} {\bibfnamefont {V.}~\bibnamefont
  {Lecl{\`e}re}}, \bibinfo {author} {\bibfnamefont {R.}~\bibnamefont {Marti}},
  \bibinfo {author} {\bibfnamefont {M.}~\bibnamefont {B{\'e}chet}}, \bibinfo
  {author} {\bibfnamefont {P.}~\bibnamefont {Fickers}}, \ and\ \bibinfo
  {author} {\bibfnamefont {P.}~\bibnamefont {Jacques}},\ }\href@noop {}
  {\bibfield  {journal} {\bibinfo  {journal} {Arch. Microbiol.}\ }\textbf
  {\bibinfo {volume} {186}},\ \bibinfo {pages} {475} (\bibinfo {year}
  {2006})}\BibitemShut {NoStop}%
\bibitem [{\citenamefont {Fauvart}\ \emph {et~al.}(2012)\citenamefont
  {Fauvart}, \citenamefont {Phillips}, \citenamefont {Bachaspatimayum},
  \citenamefont {Verstraeten}, \citenamefont {Fransaer}, \citenamefont
  {Michiels},\ and\ \citenamefont {Vermant}}]{FPB+2012sm}%
  \BibitemOpen
  \bibfield  {author} {\bibinfo {author} {\bibfnamefont {M.}~\bibnamefont
  {Fauvart}}, \bibinfo {author} {\bibfnamefont {P.}~\bibnamefont {Phillips}},
  \bibinfo {author} {\bibfnamefont {D.}~\bibnamefont {Bachaspatimayum}},
  \bibinfo {author} {\bibfnamefont {N.}~\bibnamefont {Verstraeten}}, \bibinfo
  {author} {\bibfnamefont {J.}~\bibnamefont {Fransaer}}, \bibinfo {author}
  {\bibfnamefont {J.}~\bibnamefont {Michiels}}, \ and\ \bibinfo {author}
  {\bibfnamefont {J.}~\bibnamefont {Vermant}},\ }\href {\doibase
  10.1039/C1SM06002C} {\bibfield  {journal} {\bibinfo  {journal} {Soft Matter}\
  }\textbf {\bibinfo {volume} {8}},\ \bibinfo {pages} {70} (\bibinfo {year}
  {2012})}\BibitemShut {NoStop}%
\bibitem [{\citenamefont {De~Dier}\ \emph {et~al.}(2015)\citenamefont
  {De~Dier}, \citenamefont {Fauvart}, \citenamefont {Michiels},\ and\
  \citenamefont {Vermant}}]{DFM+2015}%
  \BibitemOpen
  \bibfield  {author} {\bibinfo {author} {\bibfnamefont {R.}~\bibnamefont
  {De~Dier}}, \bibinfo {author} {\bibfnamefont {M.}~\bibnamefont {Fauvart}},
  \bibinfo {author} {\bibfnamefont {J.}~\bibnamefont {Michiels}}, \ and\
  \bibinfo {author} {\bibfnamefont {J.}~\bibnamefont {Vermant}},\ }\enquote
  {\bibinfo {title} {The role of biosurfactants in bacterial systems},}\ in\
  \href@noop {} {\emph {\bibinfo {booktitle} {The Physical Basis of Bacterial
  Quorum Communication}}},\ \bibinfo {editor} {edited by\ \bibinfo {editor}
  {\bibfnamefont {S.}~\bibnamefont {Hagen}}}\ (\bibinfo  {publisher}
  {Springer},\ \bibinfo {year} {2015})\ Chap.\ \bibinfo {chapter} {The Role of
  Biosurfactants in Bacterial Systems}, pp.\ \bibinfo {pages}
  {189--204}\BibitemShut {NoStop}%
\bibitem [{\citenamefont {Caiazza}\ \emph {et~al.}(2005)\citenamefont
  {Caiazza}, \citenamefont {Shanks},\ and\ \citenamefont
  {O'toole}}]{CSO2005job}%
  \BibitemOpen
  \bibfield  {author} {\bibinfo {author} {\bibfnamefont {N.~C.}\ \bibnamefont
  {Caiazza}}, \bibinfo {author} {\bibfnamefont {R.~M.}\ \bibnamefont {Shanks}},
  \ and\ \bibinfo {author} {\bibfnamefont {G.}~\bibnamefont {O'toole}},\
  }\href@noop {} {\bibfield  {journal} {\bibinfo  {journal} {J. Bacteriol.}\
  }\textbf {\bibinfo {volume} {187}},\ \bibinfo {pages} {7351} (\bibinfo {year}
  {2005})}\BibitemShut {NoStop}%
\bibitem [{\citenamefont {Daniels}\ \emph {et~al.}(2006)\citenamefont
  {Daniels}, \citenamefont {Reynaert}, \citenamefont {Hoekstra}, \citenamefont
  {Verreth}, \citenamefont {Janssens}, \citenamefont {Braeken}, \citenamefont
  {Fauvart}, \citenamefont {Beullens}, \citenamefont {Heusdens}, \citenamefont
  {Lambrichts} \emph {et~al.}}]{DRH+2006pnasu}%
  \BibitemOpen
  \bibfield  {author} {\bibinfo {author} {\bibfnamefont {R.}~\bibnamefont
  {Daniels}}, \bibinfo {author} {\bibfnamefont {S.}~\bibnamefont {Reynaert}},
  \bibinfo {author} {\bibfnamefont {H.}~\bibnamefont {Hoekstra}}, \bibinfo
  {author} {\bibfnamefont {C.}~\bibnamefont {Verreth}}, \bibinfo {author}
  {\bibfnamefont {J.}~\bibnamefont {Janssens}}, \bibinfo {author}
  {\bibfnamefont {K.}~\bibnamefont {Braeken}}, \bibinfo {author} {\bibfnamefont
  {M.}~\bibnamefont {Fauvart}}, \bibinfo {author} {\bibfnamefont
  {S.}~\bibnamefont {Beullens}}, \bibinfo {author} {\bibfnamefont
  {C.}~\bibnamefont {Heusdens}}, \bibinfo {author} {\bibfnamefont
  {I.}~\bibnamefont {Lambrichts}},  \emph {et~al.},\ }\href@noop {} {\bibfield
  {journal} {\bibinfo  {journal} {Proc. Natl. Acad. Sci. USA}\ }\textbf
  {\bibinfo {volume} {103}},\ \bibinfo {pages} {14965} (\bibinfo {year}
  {2006})}\BibitemShut {NoStop}%
\bibitem [{\citenamefont {Be'er}\ \emph {et~al.}(2009)\citenamefont {Be'er},
  \citenamefont {Smith}, \citenamefont {Zhang}, \citenamefont {Florin},
  \citenamefont {Payne},\ and\ \citenamefont {Swinney}}]{BSZ+2009job}%
  \BibitemOpen
  \bibfield  {author} {\bibinfo {author} {\bibfnamefont {A.}~\bibnamefont
  {Be'er}}, \bibinfo {author} {\bibfnamefont {R.~S.}\ \bibnamefont {Smith}},
  \bibinfo {author} {\bibfnamefont {H.}~\bibnamefont {Zhang}}, \bibinfo
  {author} {\bibfnamefont {E.-L.}\ \bibnamefont {Florin}}, \bibinfo {author}
  {\bibfnamefont {S.~M.}\ \bibnamefont {Payne}}, \ and\ \bibinfo {author}
  {\bibfnamefont {H.~L.}\ \bibnamefont {Swinney}},\ }\href@noop {} {\bibfield
  {journal} {\bibinfo  {journal} {J. Bacteriol.}\ }\textbf {\bibinfo {volume}
  {191}},\ \bibinfo {pages} {5758} (\bibinfo {year} {2009})}\BibitemShut
  {NoStop}%
\bibitem [{\citenamefont {Kinsinger}\ \emph {et~al.}(2003)\citenamefont
  {Kinsinger}, \citenamefont {Shirk},\ and\ \citenamefont {Fall}}]{KSF2003job}%
  \BibitemOpen
  \bibfield  {author} {\bibinfo {author} {\bibfnamefont {R.~F.}\ \bibnamefont
  {Kinsinger}}, \bibinfo {author} {\bibfnamefont {M.~C.}\ \bibnamefont
  {Shirk}}, \ and\ \bibinfo {author} {\bibfnamefont {R.}~\bibnamefont {Fall}},\
  }\href@noop {} {\bibfield  {journal} {\bibinfo  {journal} {J. Bacteriol.}\
  }\textbf {\bibinfo {volume} {185}},\ \bibinfo {pages} {5627} (\bibinfo {year}
  {2003})}\BibitemShut {NoStop}%
\bibitem [{\citenamefont {Angelini}\ \emph {et~al.}(2009)\citenamefont
  {Angelini}, \citenamefont {Roper}, \citenamefont {Kolter}, \citenamefont
  {Weitz},\ and\ \citenamefont {Brenner}}]{ARK+2009pnasu}%
  \BibitemOpen
  \bibfield  {author} {\bibinfo {author} {\bibfnamefont {T.}~\bibnamefont
  {Angelini}}, \bibinfo {author} {\bibfnamefont {M.}~\bibnamefont {Roper}},
  \bibinfo {author} {\bibfnamefont {R.}~\bibnamefont {Kolter}}, \bibinfo
  {author} {\bibfnamefont {D.~A.}\ \bibnamefont {Weitz}}, \ and\ \bibinfo
  {author} {\bibfnamefont {M.~P.}\ \bibnamefont {Brenner}},\ }\href {\doibase
  10.1002/bit.20917} {\bibfield  {journal} {\bibinfo  {journal} {Proc. Natl.
  Acad. Sci. USA}\ }\textbf {\bibinfo {volume} {106}},\ \bibinfo {pages}
  {18109} (\bibinfo {year} {2009})}\BibitemShut {NoStop}%
\bibitem [{\citenamefont {Tremblay}\ \emph {et~al.}(2007)\citenamefont
  {Tremblay}, \citenamefont {Richardson}, \citenamefont {Lépine},\ and\
  \citenamefont {Déziel}}]{TRL+2007em}%
  \BibitemOpen
  \bibfield  {author} {\bibinfo {author} {\bibfnamefont {J.}~\bibnamefont
  {Tremblay}}, \bibinfo {author} {\bibfnamefont {A.-P.}\ \bibnamefont
  {Richardson}}, \bibinfo {author} {\bibfnamefont {F.}~\bibnamefont {Lépine}},
  \ and\ \bibinfo {author} {\bibfnamefont {E.}~\bibnamefont {Déziel}},\ }\href
  {\doibase 10.1111/j.1462-2920.2007.01396.x} {\bibfield  {journal} {\bibinfo
  {journal} {Environ. Microbiol.}\ }\textbf {\bibinfo {volume} {9}},\ \bibinfo
  {pages} {2622} (\bibinfo {year} {2007})}\BibitemShut {NoStop}%
\bibitem [{\citenamefont {Matar}\ and\ \citenamefont
  {Craster}(2009)}]{MC2009sm}%
  \BibitemOpen
  \bibfield  {author} {\bibinfo {author} {\bibfnamefont {O.~K.}\ \bibnamefont
  {Matar}}\ and\ \bibinfo {author} {\bibfnamefont {R.~V.}\ \bibnamefont
  {Craster}},\ }\href {\doibase 10.1039/b908719m} {\bibfield  {journal}
  {\bibinfo  {journal} {Soft Matter}\ }\textbf {\bibinfo {volume} {5}},\
  \bibinfo {pages} {3801} (\bibinfo {year} {2009})}\BibitemShut {NoStop}%
\bibitem [{\citenamefont {Marmur}\ and\ \citenamefont
  {Lelah}(1981)}]{ML1981cec}%
  \BibitemOpen
  \bibfield  {author} {\bibinfo {author} {\bibfnamefont {A.}~\bibnamefont
  {Marmur}}\ and\ \bibinfo {author} {\bibfnamefont {M.~D.}\ \bibnamefont
  {Lelah}},\ }\href@noop {} {\bibfield  {journal} {\bibinfo  {journal} {Chem.
  Eng. Commun.}\ }\textbf {\bibinfo {volume} {13}},\ \bibinfo {pages} {133}
  (\bibinfo {year} {1981})}\BibitemShut {NoStop}%
\bibitem [{\citenamefont {Troian}\ \emph {et~al.}(1989)\citenamefont {Troian},
  \citenamefont {Wu},\ and\ \citenamefont {Safran}}]{TWS1989prl}%
  \BibitemOpen
  \bibfield  {author} {\bibinfo {author} {\bibfnamefont {S.~M.}\ \bibnamefont
  {Troian}}, \bibinfo {author} {\bibfnamefont {X.~L.}\ \bibnamefont {Wu}}, \
  and\ \bibinfo {author} {\bibfnamefont {S.~A.}\ \bibnamefont {Safran}},\
  }\href {\doibase 10.1103/PhysRevLett.62.1496} {\bibfield  {journal} {\bibinfo
   {journal} {Phys. Rev. Lett.}\ }\textbf {\bibinfo {volume} {62}},\ \bibinfo
  {pages} {1496} (\bibinfo {year} {1989})}\BibitemShut {NoStop}%
\bibitem [{\citenamefont {He}\ and\ \citenamefont
  {Ketterson}(1995)}]{HK1995pf}%
  \BibitemOpen
  \bibfield  {author} {\bibinfo {author} {\bibfnamefont {S.}~\bibnamefont
  {He}}\ and\ \bibinfo {author} {\bibfnamefont {J.}~\bibnamefont {Ketterson}},\
  }\href@noop {} {\bibfield  {journal} {\bibinfo  {journal} {Phys. Fluids}\
  }\textbf {\bibinfo {volume} {7}},\ \bibinfo {pages} {2640} (\bibinfo {year}
  {1995})}\BibitemShut {NoStop}%
\bibitem [{\citenamefont {Cachile}\ \emph {et~al.}(1999)\citenamefont
  {Cachile}, \citenamefont {Cazabat}, \citenamefont {Bardon}, \citenamefont
  {Valignat},\ and\ \citenamefont {Vandenbrouck}}]{CCB+1999csa}%
  \BibitemOpen
  \bibfield  {author} {\bibinfo {author} {\bibfnamefont {M.}~\bibnamefont
  {Cachile}}, \bibinfo {author} {\bibfnamefont {A.}~\bibnamefont {Cazabat}},
  \bibinfo {author} {\bibfnamefont {S.}~\bibnamefont {Bardon}}, \bibinfo
  {author} {\bibfnamefont {M.}~\bibnamefont {Valignat}}, \ and\ \bibinfo
  {author} {\bibfnamefont {F.}~\bibnamefont {Vandenbrouck}},\ }\href@noop {}
  {\bibfield  {journal} {\bibinfo  {journal} {Colloids Surf., A}\ }\textbf
  {\bibinfo {volume} {159}},\ \bibinfo {pages} {47} (\bibinfo {year}
  {1999})}\BibitemShut {NoStop}%
\bibitem [{\citenamefont {Afsar-Siddiqui}\ \emph {et~al.}(2003)\citenamefont
  {Afsar-Siddiqui}, \citenamefont {Luckham},\ and\ \citenamefont
  {Matar}}]{ALM2003l}%
  \BibitemOpen
  \bibfield  {author} {\bibinfo {author} {\bibfnamefont {A.~B.}\ \bibnamefont
  {Afsar-Siddiqui}}, \bibinfo {author} {\bibfnamefont {P.~F.}\ \bibnamefont
  {Luckham}}, \ and\ \bibinfo {author} {\bibfnamefont {O.~K.}\ \bibnamefont
  {Matar}},\ }\href@noop {} {\bibfield  {journal} {\bibinfo  {journal}
  {Langmuir}\ }\textbf {\bibinfo {volume} {19}},\ \bibinfo {pages} {696}
  (\bibinfo {year} {2003})}\BibitemShut {NoStop}%
\bibitem [{\citenamefont {Afsar-Siddiqui}\ \emph {et~al.}(2004)\citenamefont
  {Afsar-Siddiqui}, \citenamefont {Luckham},\ and\ \citenamefont
  {Matar}}]{ALM2004l}%
  \BibitemOpen
  \bibfield  {author} {\bibinfo {author} {\bibfnamefont {A.}~\bibnamefont
  {Afsar-Siddiqui}}, \bibinfo {author} {\bibfnamefont {P.}~\bibnamefont
  {Luckham}}, \ and\ \bibinfo {author} {\bibfnamefont {O.}~\bibnamefont
  {Matar}},\ }\href {\doibase 10.1021/la040041z} {\bibfield  {journal}
  {\bibinfo  {journal} {Langmuir}\ }\textbf {\bibinfo {volume} {20}},\ \bibinfo
  {pages} {7575} (\bibinfo {year} {2004})}\BibitemShut {NoStop}%
\bibitem [{\citenamefont {Troian}\ \emph {et~al.}(1990)\citenamefont {Troian},
  \citenamefont {Herbolzheimer},\ and\ \citenamefont {Safran}}]{THS1990prl}%
  \BibitemOpen
  \bibfield  {author} {\bibinfo {author} {\bibfnamefont {S.}~\bibnamefont
  {Troian}}, \bibinfo {author} {\bibfnamefont {E.}~\bibnamefont
  {Herbolzheimer}}, \ and\ \bibinfo {author} {\bibfnamefont {S.}~\bibnamefont
  {Safran}},\ }\href@noop {} {\bibfield  {journal} {\bibinfo  {journal} {Phys.
  Rev. Lett.}\ }\textbf {\bibinfo {volume} {65}},\ \bibinfo {pages} {333}
  (\bibinfo {year} {1990})}\BibitemShut {NoStop}%
\bibitem [{\citenamefont {Matar}\ and\ \citenamefont
  {Troian}(1999)}]{MT1999pof}%
  \BibitemOpen
  \bibfield  {author} {\bibinfo {author} {\bibfnamefont {O.~K.}\ \bibnamefont
  {Matar}}\ and\ \bibinfo {author} {\bibfnamefont {S.~M.}\ \bibnamefont
  {Troian}},\ }\href {\doibase http://dx.doi.org/10.1063/1.870185} {\bibfield
  {journal} {\bibinfo  {journal} {Phys. Fluids}\ }\textbf {\bibinfo {volume}
  {11}},\ \bibinfo {pages} {3232} (\bibinfo {year} {1999})}\BibitemShut
  {NoStop}%
\bibitem [{\citenamefont {Warner}\ \emph {et~al.}(2004)\citenamefont {Warner},
  \citenamefont {Craster},\ and\ \citenamefont {Matar}}]{WCM2004pof}%
  \BibitemOpen
  \bibfield  {author} {\bibinfo {author} {\bibfnamefont {M.}~\bibnamefont
  {Warner}}, \bibinfo {author} {\bibfnamefont {R.}~\bibnamefont {Craster}}, \
  and\ \bibinfo {author} {\bibfnamefont {O.}~\bibnamefont {Matar}},\
  }\href@noop {} {\bibfield  {journal} {\bibinfo  {journal} {Phys. Fluids}\
  }\textbf {\bibinfo {volume} {16}},\ \bibinfo {pages} {2933} (\bibinfo {year}
  {2004})}\BibitemShut {NoStop}%
\bibitem [{\citenamefont {Craster}\ and\ \citenamefont
  {Matar}(2006)}]{CM2006pof}%
  \BibitemOpen
  \bibfield  {author} {\bibinfo {author} {\bibfnamefont {R.}~\bibnamefont
  {Craster}}\ and\ \bibinfo {author} {\bibfnamefont {O.}~\bibnamefont
  {Matar}},\ }\href@noop {} {\bibfield  {journal} {\bibinfo  {journal} {Phys.
  Fluids}\ }\textbf {\bibinfo {volume} {18}},\ \bibinfo {pages} {032103}
  (\bibinfo {year} {2006})}\BibitemShut {NoStop}%
\bibitem [{\citenamefont {Marrocco}\ \emph {et~al.}(2010)\citenamefont
  {Marrocco}, \citenamefont {Henry}, \citenamefont {Holland}, \citenamefont
  {Plapp}, \citenamefont {S{\'e}ror},\ and\ \citenamefont
  {Perthame}}]{MHH+2010mmonp}%
  \BibitemOpen
  \bibfield  {author} {\bibinfo {author} {\bibfnamefont {A.}~\bibnamefont
  {Marrocco}}, \bibinfo {author} {\bibfnamefont {H.}~\bibnamefont {Henry}},
  \bibinfo {author} {\bibfnamefont {I.}~\bibnamefont {Holland}}, \bibinfo
  {author} {\bibfnamefont {M.}~\bibnamefont {Plapp}}, \bibinfo {author}
  {\bibfnamefont {S.}~\bibnamefont {S{\'e}ror}}, \ and\ \bibinfo {author}
  {\bibfnamefont {B.}~\bibnamefont {Perthame}},\ }\href@noop {} {\bibfield
  {journal} {\bibinfo  {journal} {Math Model Nat Phenom}\ }\textbf {\bibinfo
  {volume} {5}},\ \bibinfo {pages} {148} (\bibinfo {year} {2010})}\BibitemShut
  {NoStop}%
\bibitem [{\citenamefont {Thiele}\ \emph {et~al.}(2012)\citenamefont {Thiele},
  \citenamefont {Archer},\ and\ \citenamefont {Plapp}}]{TAP2012pf}%
  \BibitemOpen
  \bibfield  {author} {\bibinfo {author} {\bibfnamefont {U.}~\bibnamefont
  {Thiele}}, \bibinfo {author} {\bibfnamefont {A.~J.}\ \bibnamefont {Archer}},
  \ and\ \bibinfo {author} {\bibfnamefont {M.}~\bibnamefont {Plapp}},\ }\href
  {\doibase 10.1063/1.4758476} {\bibfield  {journal} {\bibinfo  {journal}
  {Phys. Fluids}\ }\textbf {\bibinfo {volume} {24}},\ \bibinfo {pages} {102107}
  (\bibinfo {year} {2012})},\ \bibinfo {note} {note that a term was missed in
  the variation of $F$ and a correction is contained in the appendix of
  \cite{thap2016prf}.}\BibitemShut {Stop}%
\bibitem [{\citenamefont {Trinschek}\ \emph {et~al.}(2017)\citenamefont
  {Trinschek}, \citenamefont {John}, \citenamefont {Lecuyer},\ and\
  \citenamefont {Thiele}}]{TJL+2017prl}%
  \BibitemOpen
  \bibfield  {author} {\bibinfo {author} {\bibfnamefont {S.}~\bibnamefont
  {Trinschek}}, \bibinfo {author} {\bibfnamefont {K.}~\bibnamefont {John}},
  \bibinfo {author} {\bibfnamefont {S.}~\bibnamefont {Lecuyer}}, \ and\
  \bibinfo {author} {\bibfnamefont {U.}~\bibnamefont {Thiele}},\ }\href@noop {}
  {\bibfield  {journal} {\bibinfo  {journal} {Phys. Rev. Lett.}\ }\textbf
  {\bibinfo {volume} {119}},\ \bibinfo {pages} {078003} (\bibinfo {year}
  {2017})}\BibitemShut {NoStop}%
\bibitem [{\citenamefont {Wang}\ and\ \citenamefont {Zhang}(2010)}]{WZ2010ssc}%
  \BibitemOpen
  \bibfield  {author} {\bibinfo {author} {\bibfnamefont {Q.}~\bibnamefont
  {Wang}}\ and\ \bibinfo {author} {\bibfnamefont {T.}~\bibnamefont {Zhang}},\
  }\href {\doibase http://dx.doi.org/10.1016/j.ssc.2010.01.021} {\bibfield
  {journal} {\bibinfo  {journal} {Solid State Comm.}\ }\textbf {\bibinfo
  {volume} {150}},\ \bibinfo {pages} {1009 } (\bibinfo {year}
  {2010})}\BibitemShut {NoStop}%
\bibitem [{\citenamefont {Klapper}\ and\ \citenamefont
  {Dockery}(2010)}]{KD2010sr}%
  \BibitemOpen
  \bibfield  {author} {\bibinfo {author} {\bibfnamefont {I.}~\bibnamefont
  {Klapper}}\ and\ \bibinfo {author} {\bibfnamefont {J.}~\bibnamefont
  {Dockery}},\ }\href {\doibase 10.1137/080739720} {\bibfield  {journal}
  {\bibinfo  {journal} {SIAM Rev.}\ }\textbf {\bibinfo {volume} {52}},\
  \bibinfo {pages} {221} (\bibinfo {year} {2010})}\BibitemShut {NoStop}%
\bibitem [{\citenamefont {Horn}\ and\ \citenamefont {Lackner}(2014)}]{HL2014}%
  \BibitemOpen
  \bibfield  {author} {\bibinfo {author} {\bibfnamefont {H.}~\bibnamefont
  {Horn}}\ and\ \bibinfo {author} {\bibfnamefont {S.}~\bibnamefont {Lackner}},\
  }in\ \href {\doibase 10.1007/10_2014_275} {{\selectlanguage {English}\emph
  {\bibinfo {booktitle} {Productive Biofilms}}}},\ \bibinfo {series} {Advances
  in Biochemical Engineering/Biotechnology}, Vol.\ \bibinfo {volume} {146},\
  \bibinfo {editor} {edited by\ \bibinfo {editor} {\bibfnamefont
  {K.}~\bibnamefont {Muffler}}\ and\ \bibinfo {editor} {\bibfnamefont
  {R.}~\bibnamefont {Ulber}}}\ (\bibinfo  {publisher} {Springer International
  Publishing},\ \bibinfo {year} {2014})\ pp.\ \bibinfo {pages}
  {53--76}\BibitemShut {NoStop}%
\bibitem [{\citenamefont {Picioreanu}\ and\ \citenamefont
  {Van~Loosdrecht}(2003)}]{PV2003}%
  \BibitemOpen
  \bibfield  {author} {\bibinfo {author} {\bibfnamefont {C.}~\bibnamefont
  {Picioreanu}}\ and\ \bibinfo {author} {\bibfnamefont {M.}~\bibnamefont
  {Van~Loosdrecht}},\ }\enquote {\bibinfo {title} {Biofilms in medicine,
  industry and environmental biotechnology - characteritics, analysis and
  control},}\ \ (\bibinfo  {publisher} {IWA Publishing},\ \bibinfo {year}
  {2003})\ Chap.\ \bibinfo {chapter} {Use of mathematical modelling to study
  biofilm development and morphology}, pp.\ \bibinfo {pages}
  {413--438}\BibitemShut {NoStop}%
\bibitem [{\citenamefont {Ward}\ and\ \citenamefont {King}(2012)}]{WK2012jem}%
  \BibitemOpen
  \bibfield  {author} {\bibinfo {author} {\bibfnamefont {J.}~\bibnamefont
  {Ward}}\ and\ \bibinfo {author} {\bibfnamefont {J.}~\bibnamefont {King}},\
  }\href {\doibase 10.1007/s10665-011-9490-4} {\bibfield  {journal} {\bibinfo
  {journal} {J. Eng. Math.}\ }\textbf {\bibinfo {volume} {73}},\ \bibinfo
  {pages} {71} (\bibinfo {year} {2012})}\BibitemShut {NoStop}%
\bibitem [{\citenamefont {Ward}\ \emph {et~al.}(2001)\citenamefont {Ward},
  \citenamefont {King}, \citenamefont {Koerber}, \citenamefont {Williams},
  \citenamefont {Croft},\ and\ \citenamefont {Sockett}}]{WKK+2001IJMAMB}%
  \BibitemOpen
  \bibfield  {author} {\bibinfo {author} {\bibfnamefont {J.}~\bibnamefont
  {Ward}}, \bibinfo {author} {\bibfnamefont {J.}~\bibnamefont {King}}, \bibinfo
  {author} {\bibfnamefont {A.}~\bibnamefont {Koerber}}, \bibinfo {author}
  {\bibfnamefont {P.}~\bibnamefont {Williams}}, \bibinfo {author}
  {\bibfnamefont {J.}~\bibnamefont {Croft}}, \ and\ \bibinfo {author}
  {\bibfnamefont {R.}~\bibnamefont {Sockett}},\ }\href@noop {} {\bibfield
  {journal} {\bibinfo  {journal} {IMA J. Math. Appl. Med. Biol.}\ }\textbf
  {\bibinfo {volume} {18}},\ \bibinfo {pages} {263} (\bibinfo {year}
  {2001})}\BibitemShut {NoStop}%
\bibitem [{\citenamefont {Pismen}(2006)}]{Pismen2006}%
  \BibitemOpen
  \bibfield  {author} {\bibinfo {author} {\bibfnamefont {L.~M.}\ \bibnamefont
  {Pismen}},\ }\href@noop {} {\emph {\bibinfo {title} {Patterns and interfaces
  in dissipative dynamics}}}\ (\bibinfo  {publisher} {Springer Science \&
  Business Media},\ \bibinfo {year} {2006})\BibitemShut {NoStop}%
\bibitem [{\citenamefont {Thiele}\ \emph {et~al.}(2016)\citenamefont {Thiele},
  \citenamefont {Archer},\ and\ \citenamefont {Pismen}}]{TAP2016prf}%
  \BibitemOpen
  \bibfield  {author} {\bibinfo {author} {\bibfnamefont {U.}~\bibnamefont
  {Thiele}}, \bibinfo {author} {\bibfnamefont {A.}~\bibnamefont {Archer}}, \
  and\ \bibinfo {author} {\bibfnamefont {L.}~\bibnamefont {Pismen}},\ }\href
  {\doibase 10.1103/PhysRevFluids.1.083903} {\bibfield  {journal} {\bibinfo
  {journal} {Phys. Rev. Fluids}\ }\textbf {\bibinfo {volume} {1}},\ \bibinfo
  {pages} {083903} (\bibinfo {year} {2016})}\BibitemShut {NoStop}%
\bibitem [{\citenamefont {Wilczek}\ \emph {et~al.}(2015)\citenamefont
  {Wilczek}, \citenamefont {Tewes}, \citenamefont {Gurevich}, \citenamefont
  {K{\"o}pf}, \citenamefont {Chi},\ and\ \citenamefont
  {Thiele}}]{WTG+2015mmnp}%
  \BibitemOpen
  \bibfield  {author} {\bibinfo {author} {\bibfnamefont {M.}~\bibnamefont
  {Wilczek}}, \bibinfo {author} {\bibfnamefont {W.~B.~H.}\ \bibnamefont
  {Tewes}}, \bibinfo {author} {\bibfnamefont {S.~V.}\ \bibnamefont {Gurevich}},
  \bibinfo {author} {\bibfnamefont {M.~H.}\ \bibnamefont {K{\"o}pf}}, \bibinfo
  {author} {\bibfnamefont {L.}~\bibnamefont {Chi}}, \ and\ \bibinfo {author}
  {\bibfnamefont {U.}~\bibnamefont {Thiele}},\ }\href {\doibase
  10.1051/mmnp/201510402} {\bibfield  {journal} {\bibinfo  {journal} {Math.
  Model. Nat. Phenom.}\ }\textbf {\bibinfo {volume} {10}},\ \bibinfo {pages}
  {44} (\bibinfo {year} {2015})}\BibitemShut {NoStop}%
\bibitem [{\citenamefont {Zhang}\ \emph {et~al.}(2014)\citenamefont {Zhang},
  \citenamefont {Seminara}, \citenamefont {Suaris}, \citenamefont {Brenner},
  \citenamefont {Weitz},\ and\ \citenamefont {Angelini}}]{ZSS+2014NJoP}%
  \BibitemOpen
  \bibfield  {author} {\bibinfo {author} {\bibfnamefont {W.}~\bibnamefont
  {Zhang}}, \bibinfo {author} {\bibfnamefont {A.}~\bibnamefont {Seminara}},
  \bibinfo {author} {\bibfnamefont {M.}~\bibnamefont {Suaris}}, \bibinfo
  {author} {\bibfnamefont {M.~P.}\ \bibnamefont {Brenner}}, \bibinfo {author}
  {\bibfnamefont {D.~A.}\ \bibnamefont {Weitz}}, \ and\ \bibinfo {author}
  {\bibfnamefont {T.~E.}\ \bibnamefont {Angelini}},\ }\href
  {http://stacks.iop.org/1367-2630/16/i=1/a=015028} {\bibfield  {journal}
  {\bibinfo  {journal} {New J. Phys.}\ }\textbf {\bibinfo {volume} {16}},\
  \bibinfo {pages} {015028} (\bibinfo {year} {2014})}\BibitemShut {NoStop}%
\bibitem [{\citenamefont {Dietrich}\ \emph {et~al.}(2013)\citenamefont
  {Dietrich}, \citenamefont {Okegbe}, \citenamefont {Price-Whelan},
  \citenamefont {Sakhtah}, \citenamefont {Hunter},\ and\ \citenamefont
  {Newman}}]{DOP+2013jb}%
  \BibitemOpen
  \bibfield  {author} {\bibinfo {author} {\bibfnamefont {L.}~\bibnamefont
  {Dietrich}}, \bibinfo {author} {\bibfnamefont {C.}~\bibnamefont {Okegbe}},
  \bibinfo {author} {\bibfnamefont {A.}~\bibnamefont {Price-Whelan}}, \bibinfo
  {author} {\bibfnamefont {H.}~\bibnamefont {Sakhtah}}, \bibinfo {author}
  {\bibfnamefont {R.}~\bibnamefont {Hunter}}, \ and\ \bibinfo {author}
  {\bibfnamefont {D.}~\bibnamefont {Newman}},\ }\href {\doibase
  10.1128/JB.02273-12} {\bibfield  {journal} {\bibinfo  {journal} {J.
  Bacteriol.}\ }\textbf {\bibinfo {volume} {195}},\ \bibinfo {pages} {1371}
  (\bibinfo {year} {2013})}\BibitemShut {NoStop}%
\bibitem [{\citenamefont {Doedel}\ and\ \citenamefont
  {Oldeman}(2009)}]{DO2009}%
  \BibitemOpen
  \bibfield  {author} {\bibinfo {author} {\bibfnamefont {E.~J.}\ \bibnamefont
  {Doedel}}\ and\ \bibinfo {author} {\bibfnamefont {B.~E.}\ \bibnamefont
  {Oldeman}},\ }\href@noop {} {\emph {\bibinfo {title} {AUTO07p: Continuation
  and Bifurcation Software for Ordinary Differential Equations}}},\ \bibinfo
  {organization} {Concordia University},\ \bibinfo {address} {Montreal}
  (\bibinfo {year} {2009})\BibitemShut {NoStop}%
\bibitem [{\citenamefont {Bastian}\ \emph
  {et~al.}(2008{\natexlab{a}})\citenamefont {Bastian}, \citenamefont {Blatt},
  \citenamefont {Dedner}, \citenamefont {Engwer}, \citenamefont {Kl{\"o}fkorn},
  \citenamefont {Kornhuber}, \citenamefont {Ohlberger},\ and\ \citenamefont
  {Sander}}]{BBD+2008c}%
  \BibitemOpen
  \bibfield  {author} {\bibinfo {author} {\bibfnamefont {P.}~\bibnamefont
  {Bastian}}, \bibinfo {author} {\bibfnamefont {M.}~\bibnamefont {Blatt}},
  \bibinfo {author} {\bibfnamefont {A.}~\bibnamefont {Dedner}}, \bibinfo
  {author} {\bibfnamefont {C.}~\bibnamefont {Engwer}}, \bibinfo {author}
  {\bibfnamefont {R.}~\bibnamefont {Kl{\"o}fkorn}}, \bibinfo {author}
  {\bibfnamefont {R.}~\bibnamefont {Kornhuber}}, \bibinfo {author}
  {\bibfnamefont {M.}~\bibnamefont {Ohlberger}}, \ and\ \bibinfo {author}
  {\bibfnamefont {O.}~\bibnamefont {Sander}},\ }\href@noop {} {\bibfield
  {journal} {\bibinfo  {journal} {Computing}\ }\textbf {\bibinfo {volume}
  {82}},\ \bibinfo {pages} {103} (\bibinfo {year}
  {2008}{\natexlab{a}})}\BibitemShut {NoStop}%
\bibitem [{\citenamefont {Bastian}\ \emph
  {et~al.}(2008{\natexlab{b}})\citenamefont {Bastian}, \citenamefont {Blatt},
  \citenamefont {Dedner}, \citenamefont {Engwer}, \citenamefont {Kl{\"o}fkorn},
  \citenamefont {Kornhuber}, \citenamefont {Ohlberger},\ and\ \citenamefont
  {Sander}}]{BBD+2008ca}%
  \BibitemOpen
  \bibfield  {author} {\bibinfo {author} {\bibfnamefont {P.}~\bibnamefont
  {Bastian}}, \bibinfo {author} {\bibfnamefont {M.}~\bibnamefont {Blatt}},
  \bibinfo {author} {\bibfnamefont {A.}~\bibnamefont {Dedner}}, \bibinfo
  {author} {\bibfnamefont {C.}~\bibnamefont {Engwer}}, \bibinfo {author}
  {\bibfnamefont {R.}~\bibnamefont {Kl{\"o}fkorn}}, \bibinfo {author}
  {\bibfnamefont {R.}~\bibnamefont {Kornhuber}}, \bibinfo {author}
  {\bibfnamefont {M.}~\bibnamefont {Ohlberger}}, \ and\ \bibinfo {author}
  {\bibfnamefont {O.}~\bibnamefont {Sander}},\ }\href@noop {} {\bibfield
  {journal} {\bibinfo  {journal} {Computing}\ }\textbf {\bibinfo {volume}
  {82}},\ \bibinfo {pages} {121} (\bibinfo {year}
  {2008}{\natexlab{b}})}\BibitemShut {NoStop}%
\bibitem [{\citenamefont {Eres}\ \emph {et~al.}(2000)\citenamefont {Eres},
  \citenamefont {Schwartz},\ and\ \citenamefont {Roy}}]{ESR2000pf}%
  \BibitemOpen
  \bibfield  {author} {\bibinfo {author} {\bibfnamefont {M.~H.}\ \bibnamefont
  {Eres}}, \bibinfo {author} {\bibfnamefont {L.~W.}\ \bibnamefont {Schwartz}},
  \ and\ \bibinfo {author} {\bibfnamefont {R.~V.}\ \bibnamefont {Roy}},\ }\href
  {\doibase 10.1063/1.870382} {\bibfield  {journal} {\bibinfo  {journal} {Phys.
  Fluids}\ }\textbf {\bibinfo {volume} {12}},\ \bibinfo {pages} {1278}
  (\bibinfo {year} {2000})}\BibitemShut {NoStop}%
\bibitem [{\citenamefont {Dijkstra}\ \emph {et~al.}(2014)\citenamefont
  {Dijkstra}, \citenamefont {Wubs}, \citenamefont {Cliffe}, \citenamefont
  {Doedel}, \citenamefont {Dragomirescu}, \citenamefont {Eckhardt},
  \citenamefont {Gelfgat}, \citenamefont {Hazel}, \citenamefont {Lucarini},
  \citenamefont {Salinger}, \citenamefont {Phipps}, \citenamefont
  {Sanchez-Umbria}, \citenamefont {Schuttelaars}, \citenamefont {Tuckerman},\
  and\ \citenamefont {Thiele}}]{DWC+2014ccp}%
  \BibitemOpen
  \bibfield  {author} {\bibinfo {author} {\bibfnamefont {H.~A.}\ \bibnamefont
  {Dijkstra}}, \bibinfo {author} {\bibfnamefont {F.~W.}\ \bibnamefont {Wubs}},
  \bibinfo {author} {\bibfnamefont {A.~K.}\ \bibnamefont {Cliffe}}, \bibinfo
  {author} {\bibfnamefont {E.}~\bibnamefont {Doedel}}, \bibinfo {author}
  {\bibfnamefont {I.~F.}\ \bibnamefont {Dragomirescu}}, \bibinfo {author}
  {\bibfnamefont {B.}~\bibnamefont {Eckhardt}}, \bibinfo {author}
  {\bibfnamefont {A.~Y.}\ \bibnamefont {Gelfgat}}, \bibinfo {author}
  {\bibfnamefont {A.}~\bibnamefont {Hazel}}, \bibinfo {author} {\bibfnamefont
  {V.}~\bibnamefont {Lucarini}}, \bibinfo {author} {\bibfnamefont {A.~G.}\
  \bibnamefont {Salinger}}, \bibinfo {author} {\bibfnamefont {E.~T.}\
  \bibnamefont {Phipps}}, \bibinfo {author} {\bibfnamefont {J.}~\bibnamefont
  {Sanchez-Umbria}}, \bibinfo {author} {\bibfnamefont {H.}~\bibnamefont
  {Schuttelaars}}, \bibinfo {author} {\bibfnamefont {L.~S.}\ \bibnamefont
  {Tuckerman}}, \ and\ \bibinfo {author} {\bibfnamefont {U.}~\bibnamefont
  {Thiele}},\ }\href {\doibase 10.4208/cicp.240912.180613a} {\bibfield
  {journal} {\bibinfo  {journal} {Commun. Comput. Phys.}\ }\textbf {\bibinfo
  {volume} {15}},\ \bibinfo {pages} {1} (\bibinfo {year} {2014})}\BibitemShut
  {NoStop}%
\bibitem [{\citenamefont {Kuznetsov}(2013)}]{Kuznetsov2013}%
  \BibitemOpen
  \bibfield  {author} {\bibinfo {author} {\bibfnamefont {Y.~A.}\ \bibnamefont
  {Kuznetsov}},\ }\href@noop {} {\emph {\bibinfo {title} {Elements of applied
  bifurcation theory}}},\ Vol.\ \bibinfo {volume} {112}\ (\bibinfo  {publisher}
  {Springer Science \& Business Media},\ \bibinfo {year} {2013})\BibitemShut
  {NoStop}%
\bibitem [{\citenamefont {Thiele}\ \emph {et~al.}(2013)\citenamefont {Thiele},
  \citenamefont {Todorova},\ and\ \citenamefont {Lopez}}]{TTL2013PRL}%
  \BibitemOpen
  \bibfield  {author} {\bibinfo {author} {\bibfnamefont {U.}~\bibnamefont
  {Thiele}}, \bibinfo {author} {\bibfnamefont {D.~V.}\ \bibnamefont
  {Todorova}}, \ and\ \bibinfo {author} {\bibfnamefont {H.}~\bibnamefont
  {Lopez}},\ }\href {\doibase 10.1103/PhysRevLett.111.117801} {\bibfield
  {journal} {\bibinfo  {journal} {Phys. Rev. Lett.}\ }\textbf {\bibinfo
  {volume} {111}},\ \bibinfo {pages} {117801} (\bibinfo {year}
  {2013})}\BibitemShut {NoStop}%
\bibitem [{\citenamefont {Wilczek}\ \emph {et~al.}(2016)\citenamefont
  {Wilczek}, \citenamefont {Zhu}, \citenamefont {Chi}, \citenamefont {Thiele},\
  and\ \citenamefont {Gurevich}}]{WZC+2016jopcm}%
  \BibitemOpen
  \bibfield  {author} {\bibinfo {author} {\bibfnamefont {M.}~\bibnamefont
  {Wilczek}}, \bibinfo {author} {\bibfnamefont {J.}~\bibnamefont {Zhu}},
  \bibinfo {author} {\bibfnamefont {L.}~\bibnamefont {Chi}}, \bibinfo {author}
  {\bibfnamefont {U.}~\bibnamefont {Thiele}}, \ and\ \bibinfo {author}
  {\bibfnamefont {S.~V.}\ \bibnamefont {Gurevich}},\ }\href@noop {} {\bibfield
  {journal} {\bibinfo  {journal} {J. Phys.: Condens. Matter}\ }\textbf
  {\bibinfo {volume} {29}},\ \bibinfo {pages} {014002} (\bibinfo {year}
  {2016})}\BibitemShut {NoStop}%
\bibitem [{\citenamefont {Williams}\ and\ \citenamefont
  {Jensen}(2001)}]{WJ2001ijam}%
  \BibitemOpen
  \bibfield  {author} {\bibinfo {author} {\bibfnamefont {H.}~\bibnamefont
  {Williams}}\ and\ \bibinfo {author} {\bibfnamefont {O.}~\bibnamefont
  {Jensen}},\ }\href {\doibase 10.1093/imamat/66.1.55} {\bibfield  {journal}
  {\bibinfo  {journal} {IMA J. Appl. Math.}\ }\textbf {\bibinfo {volume}
  {66}},\ \bibinfo {pages} {55} (\bibinfo {year} {2001})}\BibitemShut {NoStop}%
\bibitem [{\citenamefont {Thiele}(2015)}]{cenosTutorialLindrop}%
  \BibitemOpen
  \bibfield  {author} {\bibinfo {author} {\bibfnamefont {U.}~\bibnamefont
  {Thiele}},\ }in\ \href {http://www.uni-muenster.de/CeNoS/Lehre/Tutorials}
  {\emph {\bibinfo {booktitle} {M{\"u}nsteranian Torturials on Nonlinear
  Science: Continuation}}},\ \bibinfo {editor} {edited by\ \bibinfo {editor}
  {\bibfnamefont {U.}~\bibnamefont {Thiele}}, \bibinfo {editor} {\bibfnamefont
  {O.}~\bibnamefont {Kamps}}, \ and\ \bibinfo {editor} {\bibfnamefont {S.~V.}\
  \bibnamefont {Gurevich}}}\ (\bibinfo  {publisher} {CeNoS},\ \bibinfo
  {address} {M\"unster},\ \bibinfo {year} {2015})\ \bibinfo {edition} {1st}\
  ed.\BibitemShut {Stop}%
\bibitem [{\citenamefont {Jensen}\ and\ \citenamefont
  {Naire}(2006)}]{JN2006jofm}%
  \BibitemOpen
  \bibfield  {author} {\bibinfo {author} {\bibfnamefont {O.}~\bibnamefont
  {Jensen}}\ and\ \bibinfo {author} {\bibfnamefont {S.}~\bibnamefont {Naire}},\
  }\href@noop {} {\bibfield  {journal} {\bibinfo  {journal} {J. Fluid Mech.}\
  }\textbf {\bibinfo {volume} {554}},\ \bibinfo {pages} {5} (\bibinfo {year}
  {2006})}\BibitemShut {NoStop}%
\bibitem [{\citenamefont {Edmonstone}\ \emph {et~al.}(2004)\citenamefont
  {Edmonstone}, \citenamefont {Matar},\ and\ \citenamefont
  {Craster}}]{EMC2004joem}%
  \BibitemOpen
  \bibfield  {author} {\bibinfo {author} {\bibfnamefont {B.}~\bibnamefont
  {Edmonstone}}, \bibinfo {author} {\bibfnamefont {O.}~\bibnamefont {Matar}}, \
  and\ \bibinfo {author} {\bibfnamefont {R.}~\bibnamefont {Craster}},\
  }\href@noop {} {\bibfield  {journal} {\bibinfo  {journal} {J Eng Math}\
  }\textbf {\bibinfo {volume} {50}},\ \bibinfo {pages} {141} (\bibinfo {year}
  {2004})}\BibitemShut {NoStop}%
\bibitem [{\citenamefont {Edmonstone}\ \emph {et~al.}(2005)\citenamefont
  {Edmonstone}, \citenamefont {Matar},\ and\ \citenamefont
  {Craster}}]{EMC2005pdnp}%
  \BibitemOpen
  \bibfield  {author} {\bibinfo {author} {\bibfnamefont {B.}~\bibnamefont
  {Edmonstone}}, \bibinfo {author} {\bibfnamefont {O.}~\bibnamefont {Matar}}, \
  and\ \bibinfo {author} {\bibfnamefont {R.}~\bibnamefont {Craster}},\
  }\href@noop {} {\bibfield  {journal} {\bibinfo  {journal} {Physica D:
  Nonlinear Phenomena}\ }\textbf {\bibinfo {volume} {209}},\ \bibinfo {pages}
  {62} (\bibinfo {year} {2005})}\BibitemShut {NoStop}%
\bibitem [{\citenamefont {Goddard}\ and\ \citenamefont
  {Naire}(2015)}]{GN2015jofm}%
  \BibitemOpen
  \bibfield  {author} {\bibinfo {author} {\bibfnamefont {J.}~\bibnamefont
  {Goddard}}\ and\ \bibinfo {author} {\bibfnamefont {S.}~\bibnamefont
  {Naire}},\ }\href {\doibase 10.1017/jfm.2015.212} {\bibfield  {journal}
  {\bibinfo  {journal} {J. Fluid Mech.}\ }\textbf {\bibinfo {volume} {772}},\
  \bibinfo {pages} {535} (\bibinfo {year} {2015})}\BibitemShut {NoStop}%
\bibitem [{\citenamefont {Thiele}\ \emph {et~al.}(2018)\citenamefont {Thiele},
  \citenamefont {Snoeijer}, \citenamefont {Trinschek},\ and\ \citenamefont
  {John}}]{TST+2018apa}%
  \BibitemOpen
  \bibfield  {author} {\bibinfo {author} {\bibfnamefont {U.}~\bibnamefont
  {Thiele}}, \bibinfo {author} {\bibfnamefont {J.~H.}\ \bibnamefont
  {Snoeijer}}, \bibinfo {author} {\bibfnamefont {S.}~\bibnamefont {Trinschek}},
  \ and\ \bibinfo {author} {\bibfnamefont {K.}~\bibnamefont {John}},\
  }\href@noop {} {\bibfield  {journal} {\bibinfo  {journal} {arXiv preprint
  arXiv:1802.04042}\ } (\bibinfo {year} {2018})}\BibitemShut {NoStop}%
\bibitem [{\citenamefont {Vlamakis}\ \emph {et~al.}(2008)\citenamefont
  {Vlamakis}, \citenamefont {Aguilar}, \citenamefont {Losick},\ and\
  \citenamefont {Kolter}}]{VAL+2008gd}%
  \BibitemOpen
  \bibfield  {author} {\bibinfo {author} {\bibfnamefont {H.}~\bibnamefont
  {Vlamakis}}, \bibinfo {author} {\bibfnamefont {C.}~\bibnamefont {Aguilar}},
  \bibinfo {author} {\bibfnamefont {R.}~\bibnamefont {Losick}}, \ and\ \bibinfo
  {author} {\bibfnamefont {R.}~\bibnamefont {Kolter}},\ }\href@noop {}
  {\bibfield  {journal} {\bibinfo  {journal} {Genes \& development}\ }\textbf
  {\bibinfo {volume} {22}},\ \bibinfo {pages} {945} (\bibinfo {year}
  {2008})}\BibitemShut {NoStop}%
\end{thebibliography}
\end{document}